\documentclass[a4paper,11pt]{article}
\pdfoutput=1 
\usepackage{jheppub_mod} 
\usepackage[english]{babel}
\usepackage[utf8x]{inputenc}
\usepackage[T1]{fontenc}

\usepackage{amsmath}
\usepackage{graphicx}
\usepackage{float}
\usepackage{subfigure}
\usepackage{url}
\usepackage[colorinlistoftodos]{todonotes}
\usepackage[colorlinks=true]{hyperref}
\usepackage{comment,color}

\title{\boldmath Sommerfeld effect in freeze-in dark matter}

\author[a]{Fucheng Zhong}
\author[b]{Xinyu Wang}

\affiliation[a]{School of Physics and Astronomy, Sun Yat-sen University, Zhuhai Campus, 2 Daxue Road, Xiangzhou District, Zhuhai, P. R. China}
\affiliation[b]{Center for Advanced Quantum Studies, Department of Physics,
Beijing Normal University, Beijing 100875, China}

\emailAdd{zhongfch@mail2.sysu.edu.cn}
\emailAdd{wangxy525@mail2.sysu.edu.cn}

\abstract{
If two annihilation products of dark matter (DM) particles are non-relativistic and coupled to a light force mediator, their plane wave functions are modified due to multiple exchanges of the force mediators. This gives rise to the Sommerfeld effect (SE). We consider the attractive and repulsive force SE on the relic density in different phases of freeze-in DM. We find that in the pure freeze-in region, the attractive/repulsive force SE slightly increases/decreases DM relic density by less than $20\%$ for TeV-scale DM. In the reannihilation region, if the portal coupling $\kappa$ is sufficiently large (by comparing the portal reaction rate to the Hubble rate), DM density will reach its equilibrium, and subsequently freeze out. Compared to the case without the SE, the presence of the attractive SE leads to an enlarged cross-section. As a result, a higher equilibrium value of DM density is reached, and a lower relic density is obtained after the subsequent freeze-out. However, the repulsive SE has the opposite influence. 
In the dark sector (DS) freeze-out region, also known as the middle flat plateau or ``mesa'' in the phase diagram, the SE has a significant impact on DM relic abundance. In this region, the attractive SE suppresses DM relic density by simultaneously enlarging the cross-section of the portal and DS internal interaction. In contrast, the repulsive SE will have the opposite effect. Finally, in the usual freeze-out region, DM relic density is suppressed or enhanced by an enlarged or reduced cross-section of the portal, respectively, due to the presence of the attractive or repulsive SE. 
In summary, when considering the constraint of producing correct DM relic abundance, the inclusion of SE in the portal reaction or DS internal reaction will modify the model parameters, resulting in a band-like possible parameter space.
}


\begin{document} 
\maketitle
\flushbottom

\section{Introduction}
\label{sec:intro}

\begin{figure}[t]
        \centering
    {\begin{minipage}{0.45\linewidth}
            \centering
        \includegraphics[width=1.0\linewidth]{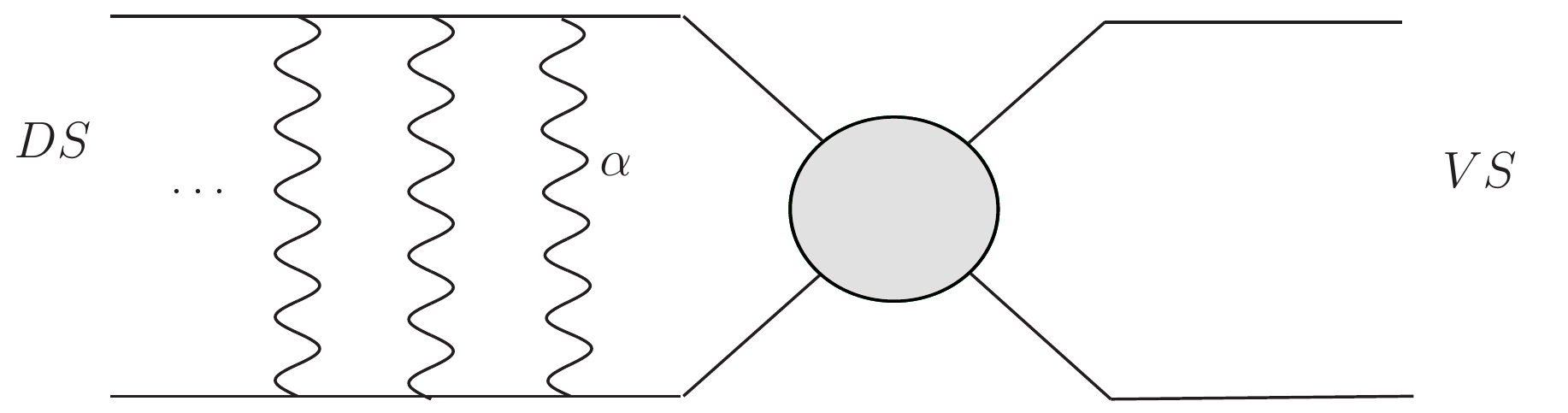}
    \end{minipage}}
    {\begin{minipage}{0.45\linewidth}
        \centering
        \includegraphics[width=1.0\linewidth]{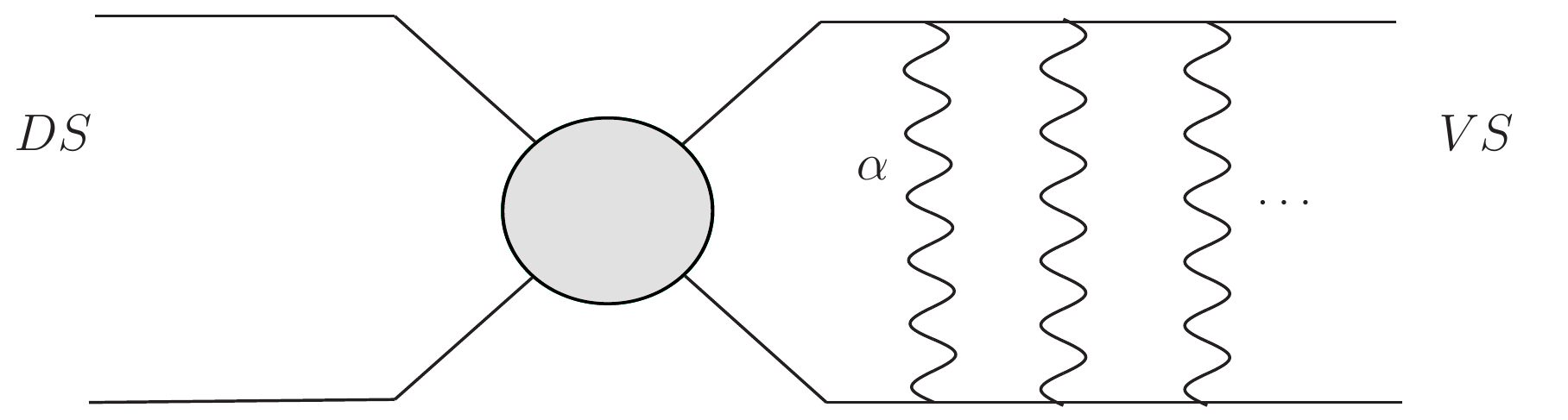}
    \end{minipage}}
    \caption{The initial SE (left panel) and the final SE (right panel). The middle circle stands for tree-level interaction.}
    \label{fig: SM enh}
\end{figure}

The Sommerfeld effect \cite{https://doi.org/10.1002/andp.19314030302}, as depicted in Fig \ref{fig: SM enh}, has been well-considered in freeze-out scenario \cite{Lee:1977ua,Kolb:1990vq} but not in the freeze-in scenario \cite{Du:2021jcj,Hall:2009bx,Bernal:2017kxu,Elahi:2014fsa,McDonald:2001vt}. We suggest the existence of the SE in the freeze-in mechanism will modify the abundance of dark matter (DM) and constraints. In the non-relativistic (low-velocity) region, if there is some long-range interaction between the annihilation particles or their annihilation products, the non-perturbative effects like the SE or bound state effects \cite{vonHarling:2014kha}, need to be considered. 

In recent times, the freeze-in dark matter mechanism has gained popularity as several novel models have proposed it as a viable explanation for the existence of dark matter. These models may include axion-like particles (ALPs), dark photons, dark Z’ bosons, Feebly Interacting Massive Particles (FIMPs), or sterile neutrinos \cite{2015PhR...555....1B}... DM was generated with an initial zero density during freeze-in. The dark sector (DS) will undergo a heating and then a subsequent cooling process. After decoupling, the DM relic density is proportional to ${\left \langle \sigma v \right \rangle}$. Hence, the mechanism requires a small coupling to match the observed relic density. The cross-section correction effect will change the relic density. If there are various particles in the DS, there will be different ``phases'' according to the various couplings between the Standard Model (SM) particles and the DS, as well as the coupling within DS. 

The relic density of WIMPs is commonly determined through the freeze-out mechanism \cite{Lee:1977ua, Kolb:1990vq}, in which DM particles are initially in thermal equilibrium with the SM and then decouple as their interaction rate becomes smaller than the Hubble expansion rate. Since the DM relic density is proportional to $ 1/{\left \langle \sigma v \right \rangle }$, a relatively large cross-section is needed to get a proper DM density. The cross-section enhancement effect will suppress the final relic density of DM, so it requires a larger mass or smaller coupling with the SM.

Non-perturbative effects, such as the SE, bound-state formation, resonance effects, and co-annihilation, have been well-studied in the context of freeze-out \cite{vonHarling:2014kha, Feng:2010zp, Arkani-Hamed:2008hhe, Hisano:2004ds, Ellis:2018jyl, Ellis:2015vna, Wang:2022avs}, but they are rare in the context of freeze-in. 
One reason for this is that freeze-out typically occurs at a low temperature compared to the mass of the DM particles ($x=m/T \sim 25$), resulting in a low velocity for the DM particles before or after annihilation, which can lead to rich non-perturbative phenomena. In contrast, at the end of freeze-in, the temperature is approximately equal to the DM mass, resulting in a relatively high temperature during DM production. At high temperatures, the annihilation cross-section becomes simpler and is roughly proportional to $1/s$, with negligible non-perturbative effects. However, we found that the SE still makes a significant contribution to the cross-section in different phases of the DM freeze-in. In this paper, we focus on the possible SE during the freeze-in process and we analyze its correction to the DM abundance. Furthermore, we provide a preliminary investigation of the SE influence on the boundary and parameter space via specific models.

We concentrate on the SE in the infrared (IR) freeze-in, rather than ultraviolet (UV) freeze-in. DM is assumed to have a negligible initial abundance, and its interaction with the particles in the bath can be so feeble\footnote{The $feeble$ coupling between DM and standard model particles is about $10^{-7}$ or less \cite{Bernal:2017kxu, Enqvist:1992va, Enqvist:2014zqa}.} that it was never in thermal equilibrium with the SM plasma. The feeble interaction leads to the continuous production of DM until the reaction rate becomes smaller than the Hubble expansion rate, then DM abundance is gradually fixed. The typical freeze-in temperature is about $x=m_r/T\sim 2-5$ \cite{Hall:2009bx}, where $m_r$ represents the relevant mass for the Boltzmann suppression. The relevant particles are moving at nearly non-relativistic velocities. Furthermore, it is natural for the relevant particles to exchange light force mediators. In the Standard Model, particle-antiparticle pairs exchange gauge bosons, and the same may happen in the dark sectors. This is the source of the SE. We suggest that the SE in freeze-in has a correction on DM abundance, which is well beyond the percent level accuracy of the observational value \cite{Planck:2018vyg}.

Our work is organized as follows: In Sec.\ref{sec: SE approx}, we make an approximation of the SE in the low-velocity limit, including the SE resonance situation, to analyze the SE for general models. Next, in Sec.\ref{sec: pure freeze-in}, we calculate the SE-corrected DM relic density and the enhancement ratio in a pure freeze-in region. We also consider DM reannihilate when the SE correction is large enough. There is a competition between the DM freeze-in and freeze-out. The SE in the DS is considered in Sec.\ref{SE in hidden}, including its effect on DS thermalization and freeze-out. It gives a similar result as the ordinary WIMP freeze-out. In Sec.\ref{sec: Model}, we provide specific models to show the SE influence on parameter space. Finally, we give some conclusions in Sec.\ref{sec:conclusion}.

\section{Sommerfeld effect and its approximations}
\label{sec: SE approx}

In this section, we focus on the contribution of the SE in different velocity regions. The low velocity-dependent part of SE often provides the most significant modifications to DM production. Identifying the contribution of different parts helps us understand how the SE modifies DM production and the physics involved.

\subsection{Approximations of Sommerfeld effects}
\begin{figure}[t]
        \centering
    {\begin{minipage}{0.45\linewidth}
            \centering
        \includegraphics[width=1.0\linewidth]{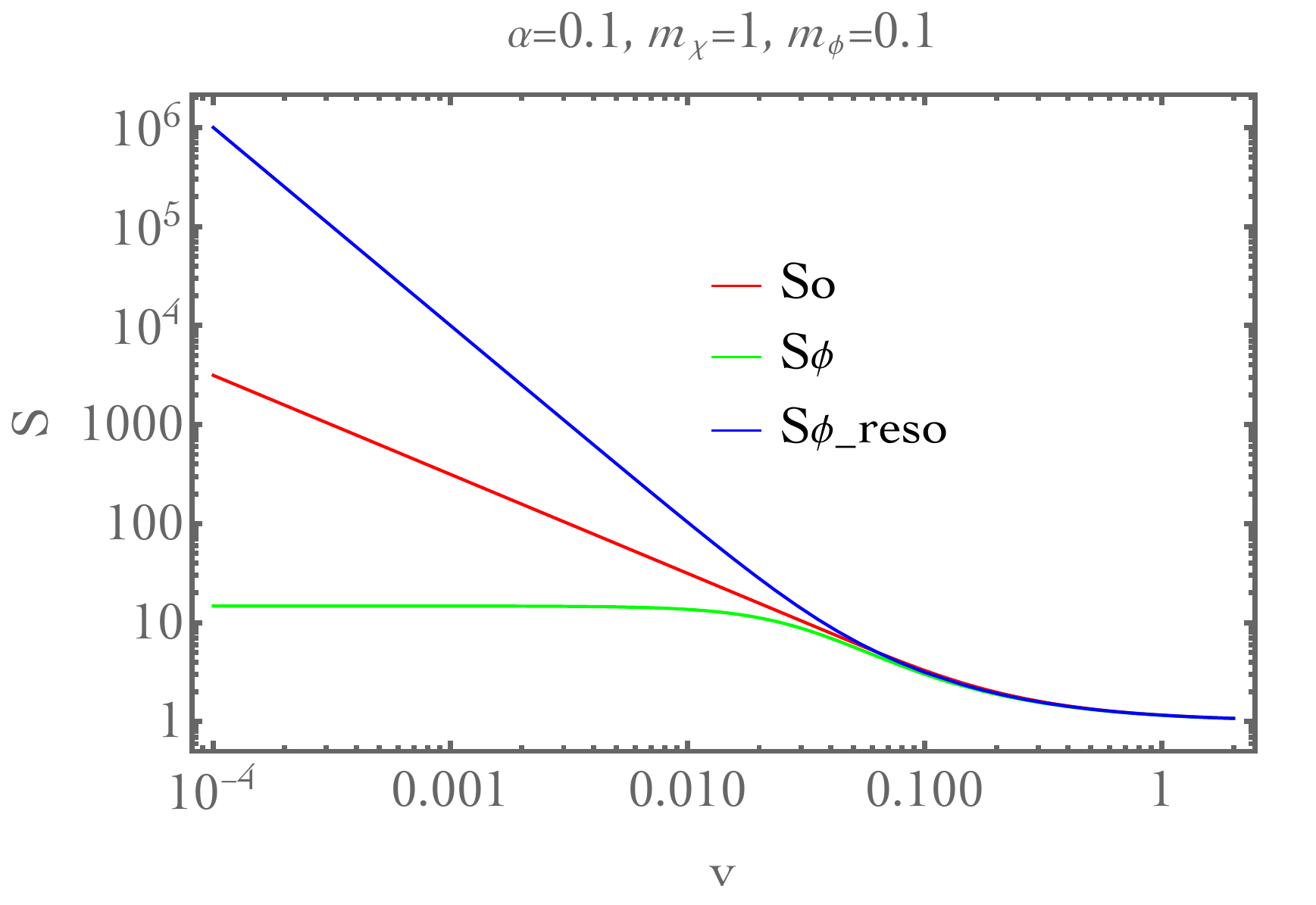}
    \end{minipage}}
    {\begin{minipage}{0.45\linewidth}
        \centering
        \includegraphics[width=1.0\linewidth]{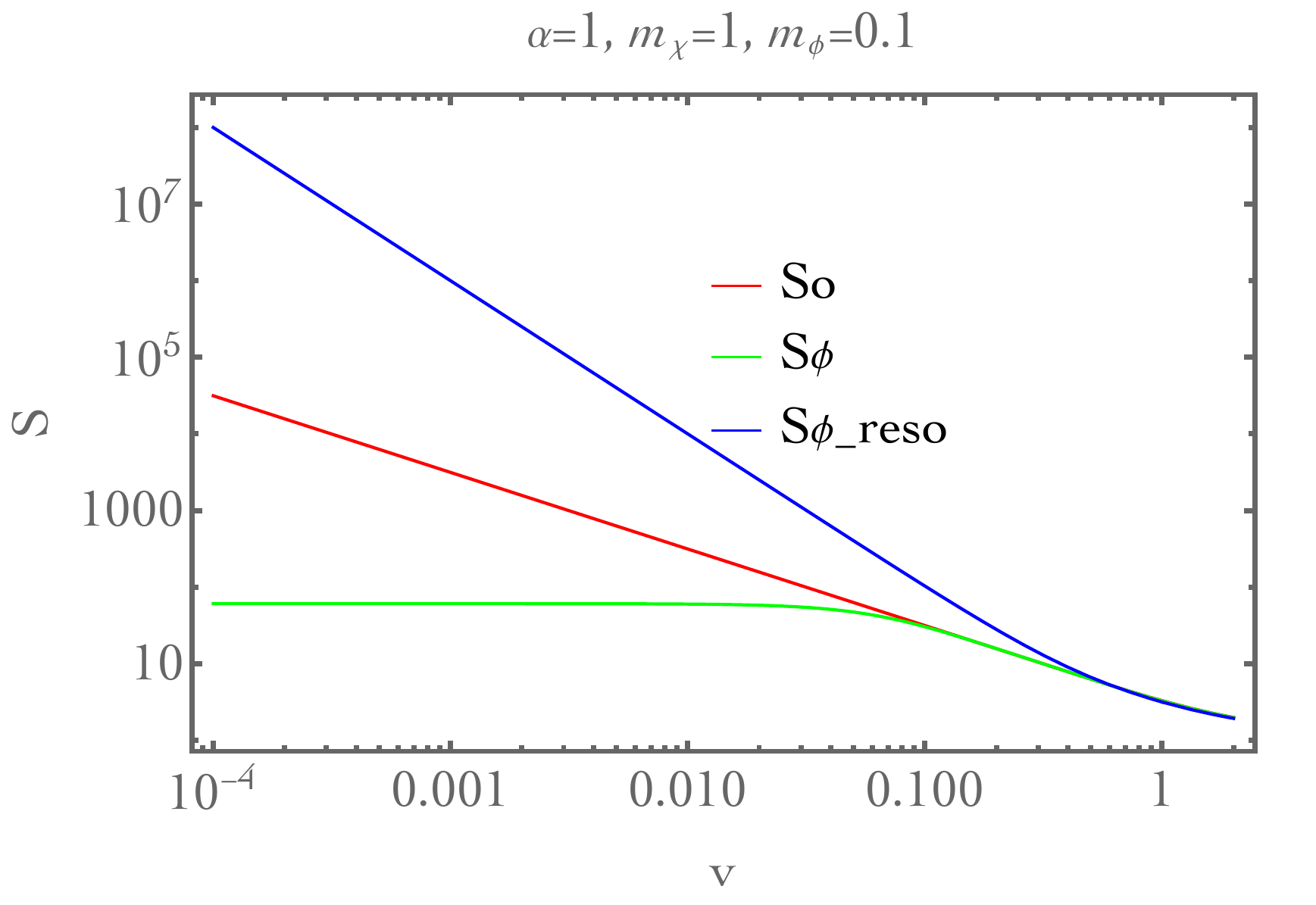}
    \end{minipage}}
    \caption{The Sommerfeld factor and its parameters are shown in the panel's title. The red, green, and blue lines correspond to the massless mediator, massive mediator, and massive mediator resonance situations, respectively. Note that for $S_o$, $m_\phi=0$; for $S_\phi\_{reso}$, $m_\phi=6\alpha m_\chi/\pi^2$. The values of $\alpha$ and $m_\chi$ remain the same in each panel.}
    \label{fig: SMF factor}
\end{figure}

The s-wave Sommerfeld factor (SF) with massless mediators is given by
\begin{align}\label{eq:SF no mass}
    S_o =  \frac{\pi \alpha /v }{1-e^{-\pi \alpha/v}},
\end{align}
where $v$ is the particle velocity in the center of mass (COM) frame and $\alpha$ is the coupling constant with the force mediators. With massive mediators, the analytic SF using the Hulthén potential approximation
\cite{2009JHEP...05..024I, Slatyer:2009vg, 2010JPhG...37j5009C, Feng:2010zp} is given by:
\begin{align} \label{eq:SF with mass}
    S_\phi = \frac{\pi}{\epsilon_v} \frac{\sinh(\frac{2 \pi \epsilon_v}{\pi^2 \epsilon_\phi/6 })}{\cosh (\frac{2 \pi \epsilon_v}{\pi^2 \epsilon_\phi/6}) - \cos (2 \pi \sqrt{\frac{1}{\pi^2 \epsilon_\phi/6}-\frac{\epsilon_v ^2}{(\pi^2 \epsilon_\phi/6)^2}})}
\end{align}
for the cases of DM singlet or zero mass gap multiplet, where $\epsilon _v = v/\alpha, \epsilon_\phi = m_\phi/(\alpha m_\chi)$. Fig. \ref{fig: SMF factor} indicates that the majority contribution of SF comes from the low-velocity region. An intuitive explanation is that annihilating particles have sufficient time to exchange many mediators, and the contribution of high-order loop diagrams cannot be neglected.

The behavior of SF under different velocities can be simplified into three parts: low-velocity resonance region where $S \sim v^{-2}$, low velocity without resonance region where $S \sim v^{-1}$, and the constant region. The SF can be simplified as:
\begin{align}
    S \simeq  a + b v^{-1} + c v^{-2},
\end{align}
where $a$ stands for the constant term, $b$ represents the part without resonance, and $c$ is for the part with resonance. Including the saturation case for the very low velocity (a constant term), the asymptotic behavior of SF under different velocity regions, DM mass, and coupling constant can be expressed as follows:
\begin{align}
S \sim  \left\{\begin{array}{lll}
            & 6 \csc^2 (\sqrt{\frac{6}{\epsilon_\phi}}) /\epsilon_\phi & \ v< \alpha^4 \epsilon_\phi \\
            & \alpha^2 / v^2 &  \  \epsilon_v << \epsilon_\phi,  \\
            & \pi \alpha / v    &  \ \epsilon_v >> \epsilon_\phi ,  v<v_{lim} \\
            & 1 & \ v_{lim}<v
            \end{array}\right.
\end{align}
In the saturation region, SF saturates at $v \sim \alpha^4 \epsilon_\phi$. The saturation is caused by the finite lifetime of bound states \cite{Hisano:2004ds, Elor:2018xku, Hisano:2003ec}. In the relativistic region, $S \sim 1$ \cite{Slatyer:2009vg, 2017JCAP...08..003K}, where $v_{lim}$ is the velocity beyond which the SE becomes negligible. Resonance occurs when there are zero-energy bound states in two body systems \cite{Arkani-Hamed:2008hhe}. The situations $\epsilon_\phi>1$ or $\epsilon_v > 1$ will lead to SF $\sim \mathcal{O}(1)$ \cite{Slatyer:2009vg}. The first situation represents a mediator with a large mass and short force range that cannot provide enough SE enhancement, and the second situation indicates that the SE is insignificant in the relativistic region.
The resonance condition can be written as:
\begin{align}
    m_\phi \simeq \frac{6 \alpha m_\chi}{\pi^2 n^2}, \quad n=1,2,3 \ldots
\end{align}
We take $n=1$ to achieve maximum enhancement.

The $p$-wave SF can be obtained through the $s$-wave SF using the following formula \cite{2017JCAP...08..003K}: 
\begin{align}
    S_{p} & = \frac{(c-1)^{2}+4 a^{2} c^{2}}{1+4 a^{2} c^{2}} \times S 
\end{align}
where $a=\epsilon_v$, $c = 6/ \pi^2 \epsilon_\phi$. 
The cross-section is contributed by different partial waves and can be expressed as $\sigma v_0=a_s+b_s v^2+b_p v^2$, where $a_s+b_s v^2$ represents the contribution from the $s$-wave, and $b_p v^2$ represents the contribution from the $p$-wave \cite{2019PhRvD..99c6016T}. Then, the total SE-modified cross-section can be presented as
\begin{align} 
    & (\sigma v)_{md} = (a_s+b_s v^2) S + (b_p v^2) S_p.
\end{align}
The majority contribution of the SE is from the $s$-wave, except under certain special conditions such as angular momentum conservation and selection rules \cite{2019PhRvD..99c6016T, Wang:2022avs}. For simplicity, we only consider the contribution from the $s$-wave in this paper. The $s$-wave SE-modified cross-section incorporating is denoted as:
\begin{align} \label{eq: enhancement sigma}
    \sigma_{i}(\tilde{s}) = \sigma(\tilde{s}) \cdot S_i, 
\end{align}
where $\tilde{s} \equiv s/4m_\chi^2$ , with $s$ represents the invariant mass and $m_\chi$ represents DM mass.

\subsection{The thermal average of the SE-modified cross-section}

The evolution of the number density of particles in the early universe is governed by the Boltzmann equation. An important term in the Boltzmann equation is the thermally averaged cross-section, which can be cast as: 
\begin{align}
    \left\langle\sigma v _{12 \rightarrow 34}\right\rangle=\frac{C}{2 T K_{2}\left(m_{1} / T\right) K_{2}\left(m_{2} / T\right)} \int_{s_{\min }}^{\infty} \sigma(s) \frac{F\left(m_{1}, m_{2}, s\right)^{2}}{m_{1}^{2} m_{2}^{2} \sqrt{s}} K_{1}(\sqrt{s} / T) \mathrm{d} s \ ,
\end{align}
where $F\left(m_{1}, m_{2}, s\right)=\frac{1}{2}\sqrt{[s-(m_1^2+m_2^2)]^2-4m_1^2m_2^2}$, and $C$ is a statistical factor. 

Assuming that DM is generated by the $f f \to \chi \chi$ reaction and 
the particle $f$ has a Boltzmann-distributed phase-space distribution, the thermally averaged cross-section can be simplified as follows \cite{1991NuPhB.360..145G, Bringmann:2021sth}:
\begin{align}
    \langle\sigma v\rangle_{i} = \frac{4x}{K_{2}{ }(x)^{2}} \int_{\tilde{s}_{min}}^{\infty} d \tilde{s} \sqrt{\tilde{s}}(\tilde{s}-1) K_{1}(2 \sqrt{\tilde{s}} x) \ \sigma_{i}(\tilde{s}),
\end{align}
where $ \tilde{s}_{min} = \max[1,(m_f/m_\chi)^2] $ represents the kinematically allowed velocity, introduced to account for the forbidden case \cite{DAgnolo:2015ujb} or final-state SE \cite{Cui:2020ppc, Wang:2022avs}. 

\section{Freeze-in with the SE and upper bound}
\label{sec: pure freeze-in}

As WIMPs are becoming increasingly constrained by experiments, alternative DM candidates with lower mass or very weak interactions have become more and more popular. For instance, ALPs, FIMPs, dark photons, and dark Higgs \cite{McDonald:2001vt, Biswas:2022vkq, Bernal:2017kxu, Chu:2011be, Hall:2009bx, Bharucha:2022lty, Berbig:2022nre, Luo:2020fdt, DEramo:2020gpr, Berbig:2022nre, Matsui:2015maa, Bae:2017dpt}... They are produced by a new mechanism called freeze-in. The difference from freeze-out is that its interaction can be very feeble, with zero initial density of the DM.



The freeze-in mechanism can be achieved in two temperature regions, UV-dominated and IR-dominated \cite{DEramo:2017ecx}. The UV-dominated region is mainly associated with non-renormalized interactions, which are suppressed at high energy scales. It seems unlikely for the SE to occur in this region. In an IR-dominated region, the SE can potentially occur at the decoupling temperature. 

Generally, the parameter space of freeze-in can be separated into four different ``phases'': pure freeze-in, DM reannihilation, DS freeze-out, and usual freeze-out, or even more (sequential freeze-in) \cite{Hall:2009bx, Hambye:2019dwd, Belanger:2020npe, Bharucha:2022lty}. The evolution of the number density of unknown DS particles is governed by the Boltzmann equations. The intractability arises from the DS particles' initial density and the non-equilibrium between DS and SM (depending on the coupling constant strength/transfer rate).

\subsection{The SE on pure freeze-in relic density}
\label{sec: Maximum SE enhancement}

The simplest form of DS contains two kinds of particles: DM particles and corresponding mediators. So we simplify the process by considering only one kind of DM particle, $\chi$, and one kind of corresponding light mediator. The differences in initial DM number density, mass, and coupling between the DS and the SM can give rise to rich phenomena in freeze-in. 
Considering a typical annihilation $ f \bar{f} \to \chi \bar{\chi}$ with a SE-corrected cross-section, the Boltzmann equation governing the DM number density is expressed as follows:
\begin{align}\label{eq: Bolz}
    \frac{dY_\chi}{dx}  & = \frac{xs}{H_{m_\chi}}(1+\frac{T}{3g_{*s}}\frac{dg_{*s}}{dT}) \cdot \left \langle \sigma v _{\chi \to f } \right  \rangle \cdot (Y^{eq \ 2}_{\chi} - Y_\chi^2) 
\end{align}
where $x=m_\chi/T, \ H_{m_\chi}= \sqrt{\frac{4 \pi^3  g_*}{45}} \frac{m_\chi^2}{m_{pl}}$. The right-hand side (RHS)  of the equation includes any reaction terms that lead to the changing number density of $\chi$. In freeze-in, $\chi$ has a feeble interaction with $f$, and its initial number density is nearly zero.
Assuming $\chi$ never reaches thermal equilibrium during the generation process (depending on the coupling $\alpha$ and number density $n_\chi$ or the portal coupling $\kappa$, we use ratio $\langle \sigma v \rangle / H$ as a criterion in Sec \ref{sec: simple 2to2 sigma} and Sec \ref{sec. reannihilation} to check whether it is in a visible sector thermal bath), as a result, the $Y_\chi^2$ term can be eliminated in equations ($Y_\chi \to 0$), and makes $g_{*s}$ constant. The $yield$ of $\chi$ can be immediately obtained [Appendix \ref{app: upper bound}] as follows:
\begin{align} \label{eq: simplest Y}
    Y_{\chi i} = C_1 \int_{\tilde{s}_{min}}^{\infty} d \tilde{s} \ \frac{\tilde{s}-1}{\tilde{s}^{3/2}} \cdot \sigma_i (\tilde{s}).
\end{align}
 Here we assume that the temperature effects will not affect $\sigma({\tilde{s}})$. A more precise analysis needs to take into account the vertex correction and thermal mass effect \cite{ Cirelli:2007xd, RevModPhys.53.43}, but some studies \cite{Binder:2018znk, 2022arXiv220903932B, Wang:2022avs} have shown the contribution is small. The $dg_{*s}/dT$ term can be neglected since it is suppressed by $\mathcal{O}(10^2)$ unless there is something like instant thermalizing or cooling (like phase transition). The ratio 
 \begin{align} \label{eq: ratio eta}
     \eta_i =  Y_{\chi i} /Y_{\chi} = \frac{\int_{\tilde{s}_{min}}^{\infty} d \tilde{s} \ \frac{\tilde{s}-1}{\tilde{s}^{3/2}} \cdot \sigma_i (\tilde{s})}{\int_{\tilde{s}_{min}}^{\infty} d \tilde{s} \ \frac{\tilde{s}-1}{\tilde{s}^{3/2}} \cdot \sigma (\tilde{s})}
 \end{align}
 gives the SE enhancement ratio (where $Y_{\chi}$ without additional indices represents the situation without the SE).

 \subsubsection{Initial State SE}
 \label{sec: The Initial State SE}
 The Initial State SE is attributed to the cross-section modified by the coupling $\alpha$ within the DS particles, illustrated in the left panel of Fig \ref{fig: SM enh}. There are two cases based on different DM masses: the kinematically forbidden case ($m_\chi < m_f$) and the usual case ($m_\chi > m_f$). We integrate over the kinematically allowed region, and the enhancement ratio is as follows:
\begin{align}
     \eta_i =  \frac{\int_{z_0}^{\infty} d \tilde{s} \ \frac{\tilde{s}-1}{\tilde{s}^{3/2}} \cdot \sigma_i (\tilde{s})}{\int_{z_0}^{\infty} d \tilde{s} \ \frac{\tilde{s}-1}{\tilde{s}^{3/2}} \cdot \sigma(\tilde{s})}.
\end{align}
where $z_0 \equiv Max[1,m_f^2/m_\chi^2]$. It is worth noting that for the convenience of freeze-out calculation, the true reaction direction should be $f \to \chi$, but the calculated cross-section is $\chi \to f$ as Eq \ref{eq: Bolz} shows. In the $m_\chi > m_f$ case, $\eta_i$ is larger because the SE is more efficient at low velocities.

\subsubsection{Final State SE}
The Final State SE \cite{Cui:2020ppc} is caused by the coupling between the SM particles, which modifies the annihilation cross-section. The right panel of Fig. \ref{fig: SM enh} depicts the situation of the Final State SE, where the coupling $\alpha$ arises from the interactions within the visible sector (VS). The Final State SE relies on the velocity of the final state particles. The formula is the same as Eq. \ref{eq: ratio eta}, except that the velocity in $S_i$ should be replaced by the final state particles ($f$ in the section \ref{sec: The Initial State SE}) velocity $v = \sqrt{1-z \tilde{s}^{-1}}$, where $z \equiv m_f^2/m_\chi^2$.

\subsubsection{Simple $2 \to 2$ cross-section}
\label{sec: simple 2to2 sigma}
The form of the $s$-wave $2 \to 2$ cross-section is simple: it approaches $ 1/\tilde{s}$ as the energy increase due to unitarity. We parameterized the cross-section by $\tilde{s}$, which is the Mandelstam variable nondimensionalized by $4m_\chi^2$, as described in [Appendix \ref{app: Kinematics}]
 \begin{align} \label{eq: 1/s type}
    \sigma_{\chi \to f}(\tilde{s}) = \frac{\kappa ^2}{4m_\chi^2 \tilde{s}} \frac{\sqrt{\tilde{s}-z}}{\sqrt{\tilde{s}-1}} \ \Theta(\tilde{s}-1)\Theta(\tilde{s}-z),
\end{align}
where $\kappa$ is the connecting/portal coupling between SM and DS; $\Theta$ is the Heaviside function. 

\begin{figure}[t]
    \centering
    {\begin{minipage}{0.45\linewidth}
        \centering
        \includegraphics[width=1.0\linewidth]{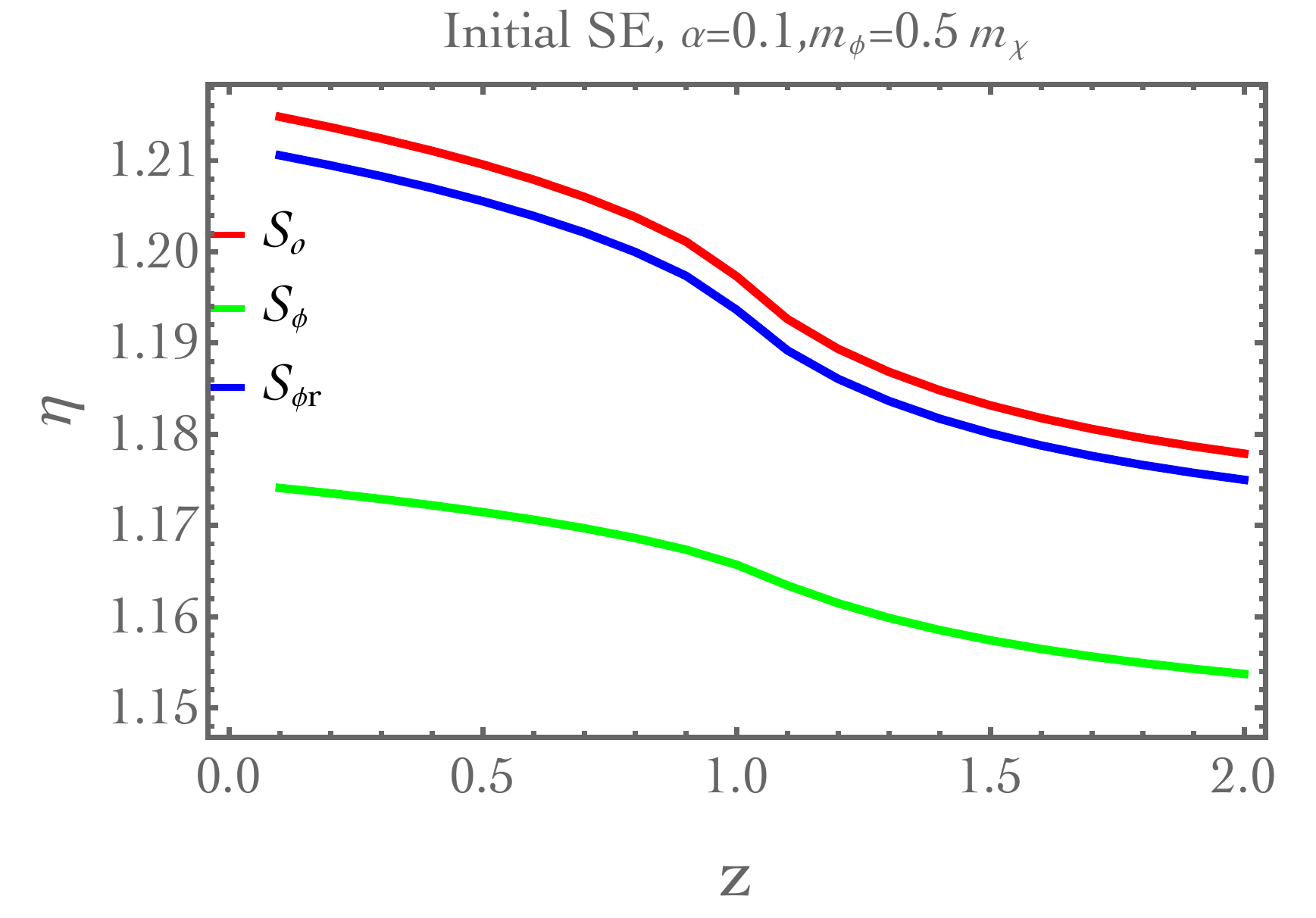}
    \end{minipage}}
    {\begin{minipage}{0.45\linewidth}
        \centering
        \includegraphics[width=1.0\linewidth]{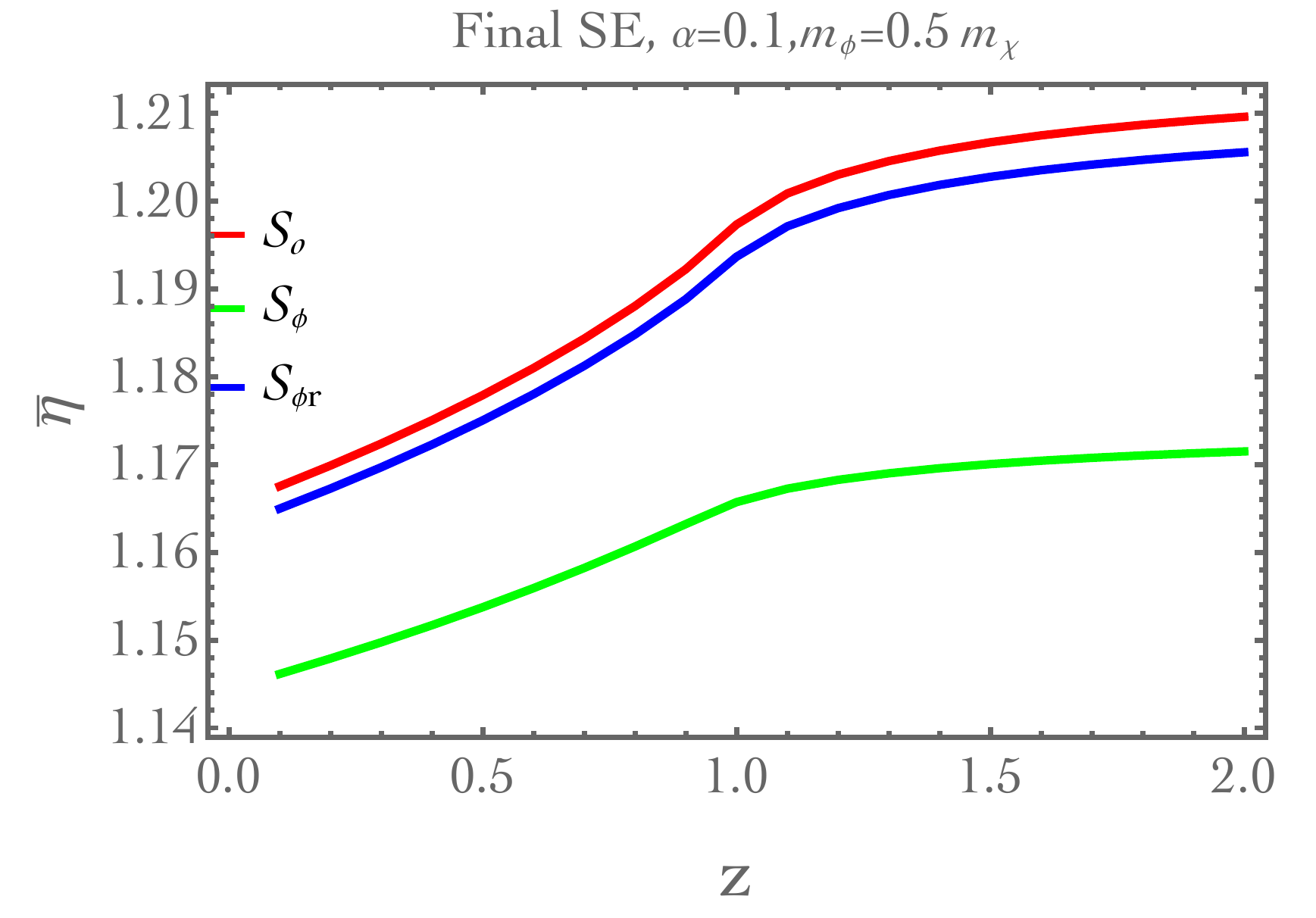}
    \end{minipage}}
    {\begin{minipage}{0.45\linewidth}
        \centering
        \includegraphics[width=1.0\linewidth]{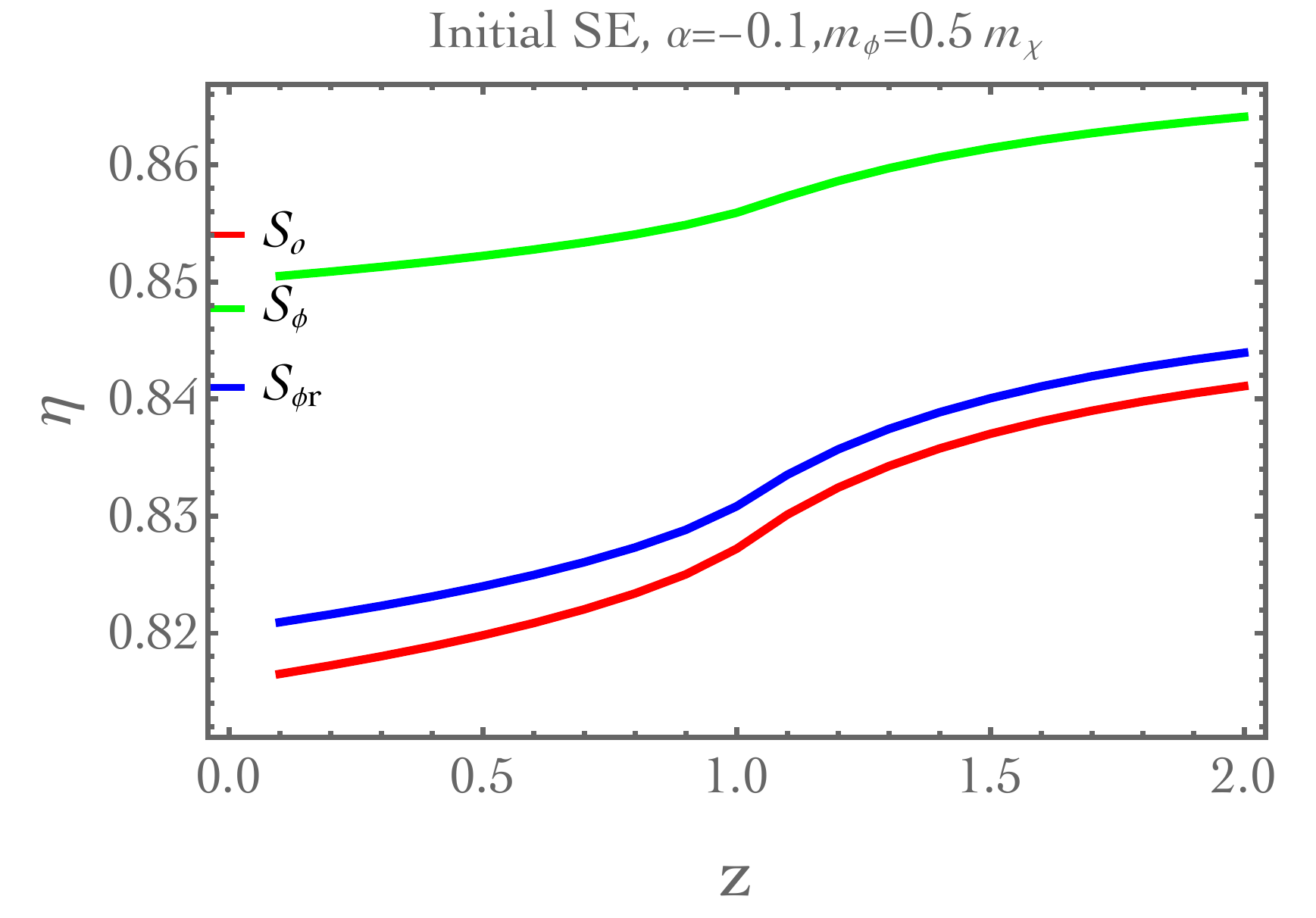}
    \end{minipage}}
    {\begin{minipage}{0.45\linewidth}
        \centering
        \includegraphics[width=1.0\linewidth]{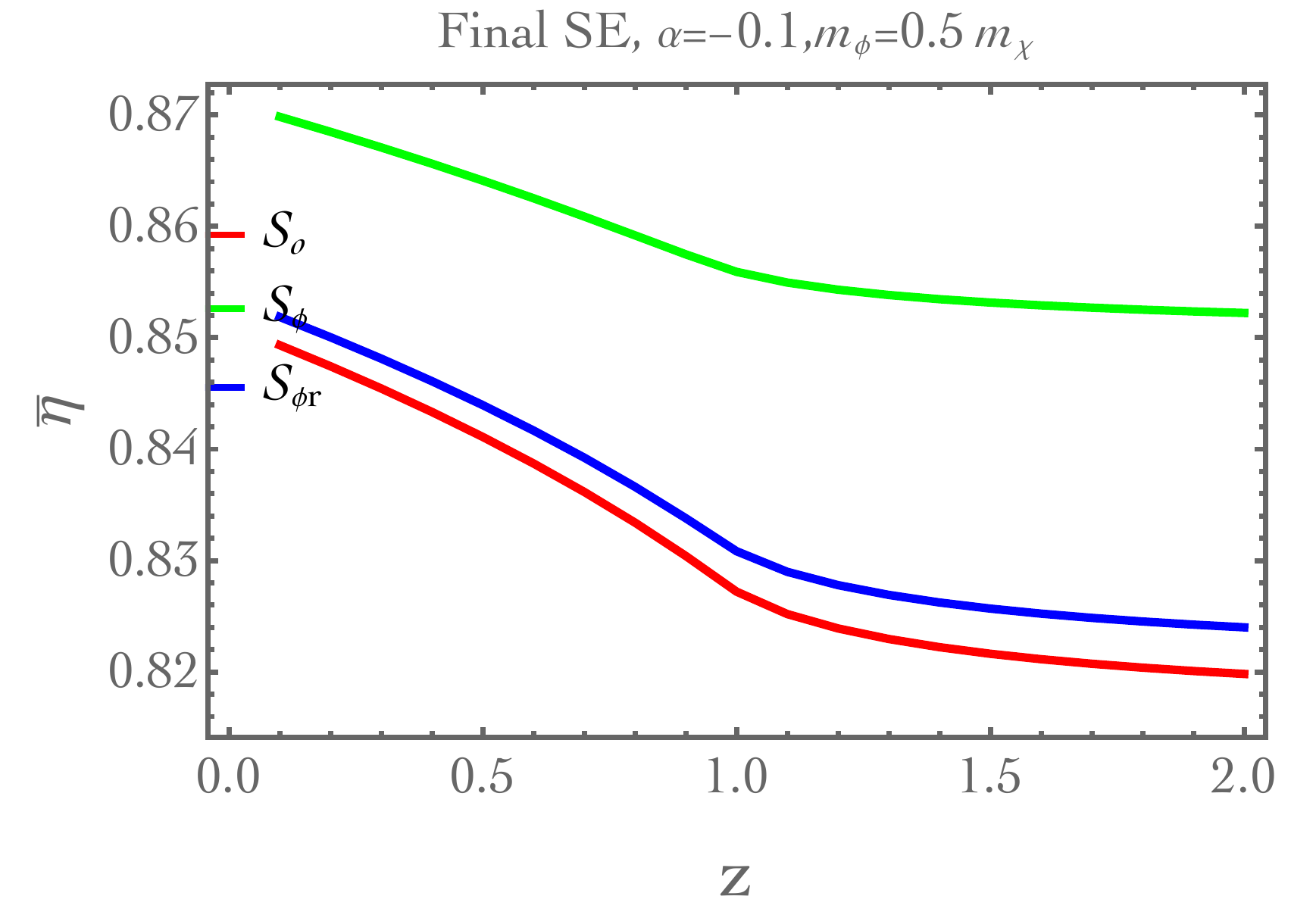}
    \end{minipage}}
    \caption{$S_o,S_\phi, S_{\phi r}$ stand for the cases of massless mediator, $0.5m_\chi$ massive mediator and resonant mediator, respectively. The panels show the SE enhancement ratio of the 4-point interaction (including the attractive and repulsive force mediator) for different $z$. Note that for $S_{\phi r}$ curve, $m_\phi$ is chosen based on the resonance parameters. The left panels are the initial SE, and the right panels are the final SE ($\bar \eta$). The top two panels depict the attractive force SE, while the bottom two panels depict the repulsive force SE.}
    \label{fig: SE for 1/s}
\end{figure}

The initial SE correction cross-section is caused by the internal DS interaction $\alpha$. To ensure that $\chi$ particle and its mediator $\phi$ are not in the thermal equilibrium, the effective cross-section should be smaller than the Hubble rate
\begin{align}
    & \left \langle \sigma v _{eff} \right  \rangle  n^{eq}_\chi < H \\
    & \left \langle \sigma v _{eff} \right  \rangle = \sqrt{ \left \langle \sigma v _{f \to \chi} \right  \rangle \left \langle \sigma v _{\chi \to \phi} \right  \rangle }.
\end{align}
Under some typical values of pure freeze-in [see Appendix \ref{app: alpha threshold}], the estimated maximum $\alpha_{th}$ is given by:
\begin{align}
    \alpha_{th} \sim \sqrt{\frac{g_*}{5 \pi}} \frac{320 \ m_\chi}{\kappa \ m_{pl}}.
\end{align}
Taking $m_\chi = 1$ TeV and $\kappa \sim 10^{-12}$, we can calculated $\alpha_{th} \sim 0.07$.
Assuming that the maximum possible coupling is $\alpha = 0.1$, the enhancement ratios of the initial SE and final SE for the 4-point interaction are shown in Figure \ref{fig: SE for 1/s}. We suggest that these are the bounds that SE can provide for the Eq. \ref{eq: 1/s type} type cross-section. The enhancement ratio for the pure freeze-in relic density is below $\mathcal{O}(1)$ for $TeV$ scale DM (about 20\% changing for $1TeV$ DM). Due to the occurrence of pure freeze-in at relatively high temperatures, it cannot provide a large cross-section correction like in freeze-out scenarios. In section \ref{sec: Model}, we will demonstrate the modification of the parameter space by the SE.

\subsection{SE in DM reannihilation}
\label{sec. reannihilation}

In the last section, $Y_\chi$ in Eq. \ref{eq: Bolz} was equal to zero due to an initially null DM density and feeble interaction. This led to a constraint on the enhancement/suppression due to the SE. Considering the case of $Y_\chi \ \ne 0 $, the term of $Y_\chi$ and its initial condition can be as follows:
\begin{align}
    \left\{Y_\chi = 0\right\} \to \left\{\begin{array}{cc}
         & Y_\chi/Y_\chi ^{eq} = \delta \\
         & Y_\chi(0) = 0
            \end{array}\right\}.
\end{align}
This means that parts of the generated DM will annihilate back to $f$. It will not have a significant influence if DM density is very small and the interaction between DM and SM is feeble enough. However, it still decreases the final number density of $\chi$. Now Eq. \ref{eq: Bolz} is rewritten as:
\begin{align}\label{eq: linear approx}
    & Y_{\chi i}  =  \int_{0}^{\infty}  dx \frac{xs}{H_{m_\chi}} \cdot \left \langle \sigma v _{ \chi \to f} \right  \rangle_i \cdot (Y^{eq }_{\chi})^2 \cdot  (1-\delta^2).
\end{align}
The $\chi$ particle is in the DS thermal bath with temperature $T'$, which satisfies $T' \ll T$ at the time of $x=0$. Here, $T'$ is a function of $T$, the temperature of the VS bath. $\chi$ is not in thermal equilibrium with the VS bath, which requires that the rate of reaction $\chi \leftrightarrow  f$ should be smaller than the Hubble rate before decoupling \cite{Kolb:1990vq}. This is denoted as $\Gamma_{connect} <  H(T)$. The reaction rate between DS and VS, $\Gamma_{connect}$, includes the forward transfer rate $\overrightarrow{ \Gamma}_{connect} $ from the VS to the DS, and the backward $\overleftarrow{ \Gamma}_{connect}$ from the DS to the VS. When the VS and DS are both in the same equilibrium, it holds that $\overrightarrow{ \Gamma}_{connect} = \overleftarrow{ \Gamma}_{connect}$. However, in the freeze-in process, the forward transfer is larger than the Hubble rate, while the backward transfer rate is smaller than the Hubble rate before decoupling ($T > T_{freeze-in}$):
\begin{align} \label{eq:freeze-in condition1}
    &  \overrightarrow{ \Gamma}_{connect}  = \left \langle \sigma v _{f \to \chi } \right  \rangle n_f^{eq} > H(T) \\ \label{eq:freeze-in condition2}
    & \overleftarrow{ \Gamma}_{connect} = \left \langle \sigma v _{\chi \to f } \right  \rangle n_\chi < H(T).
\end{align}
The above conditions ensure that the DM relic density monotonically increases, but an interesting thing happens if condition \ref{eq:freeze-in condition2} inverts as $\chi$ number density increases or the cross-section is enhanced in low temperature due to the SE. This leads to the subsequent thermalization of $\chi$ by the VS bath, followed by $\chi$ freeze-out. If the backward transfer rate is larger than Hubble, the freeze-in generated DM will be heated, then cools down, and finally freeze out.

Assuming DS is at temperature $T'$, which is proportional to $T$, the relationship is as follows:
\begin{align}
    T'/T = \xi(T).
\end{align}
$\chi$ is a radiation or matter density configuration: 
\begin{align}
    & \delta = (x/x')^3 =  \xi^3 , \ \text{radiation }\chi\\
    & \delta \simeq  \frac{ (x \xi)^{3/2} }{e^{-x'}}, \ \text{ matter }\chi.
\end{align}
Ensuring the freeze-in conditions Eq. \ref{eq:freeze-in condition1}, \ref{eq:freeze-in condition2}, one can get the constraints of $\xi$ [Appendix \ref{app: balance condition}]:
\begin{align}
    & 0 < \xi < 1 \  \ \ \text{radiation }\chi \\ 
    & \xi = 0  \ \ \ \text{matter }\chi.
\end{align}
A more restrictive constraint on $\xi$ for radiation $\chi$ was given in \cite{Chu:2011be}. If the DS bath configuration is dominated by matter or radiation with $T' << T$, then $\xi \to 0$ and $\delta^2 \to 0$, which corresponds to the simplest SE situation shown in Sec. \ref{sec: Maximum SE enhancement}.

\begin{figure}[t]
    \centering
    {\begin{minipage}{0.45\linewidth}
        \centering
        \includegraphics[width=1.0\linewidth]{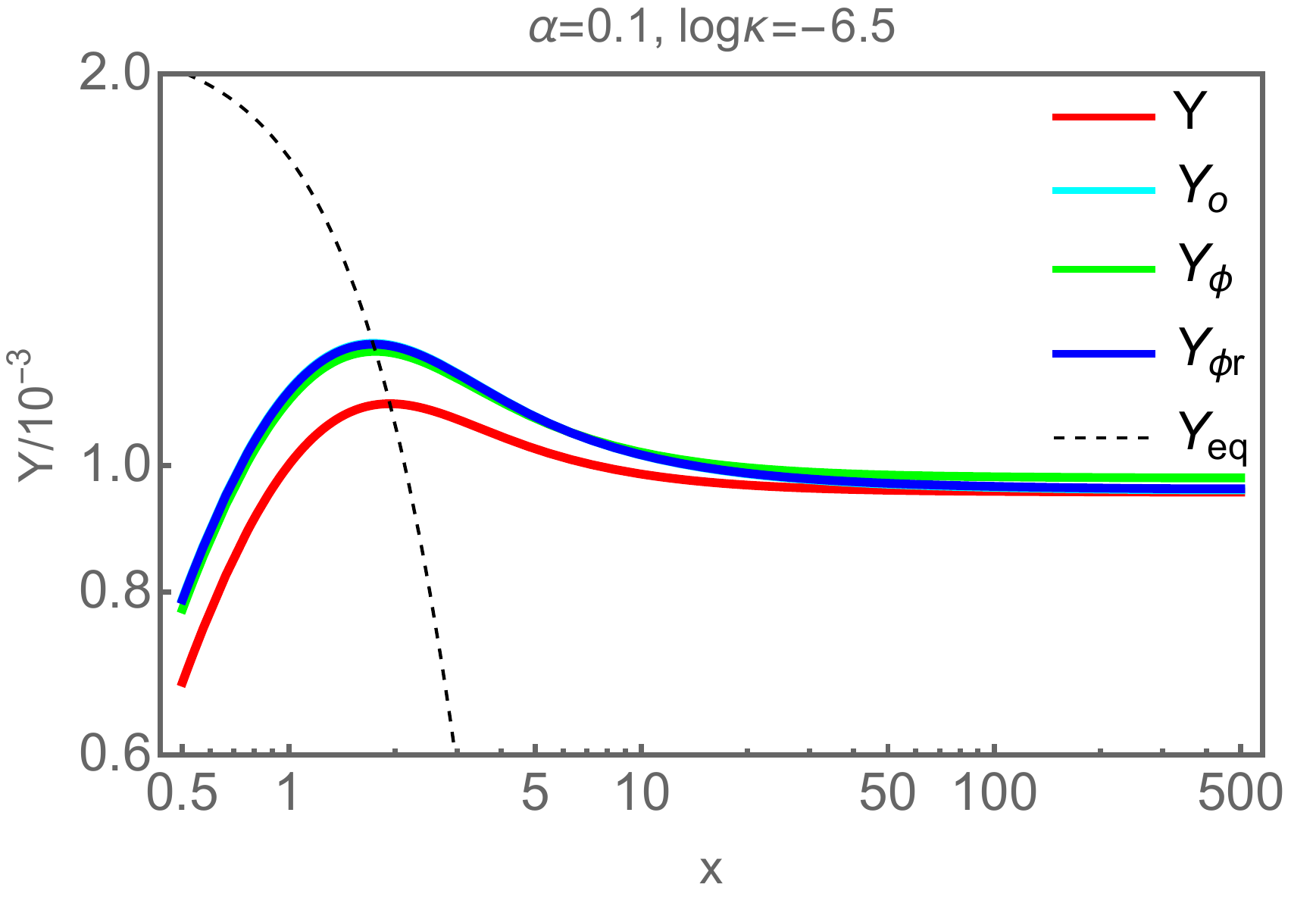}
    \end{minipage}}
    {\begin{minipage}{0.45\linewidth}
        \centering
        \includegraphics[width=1.0\linewidth]{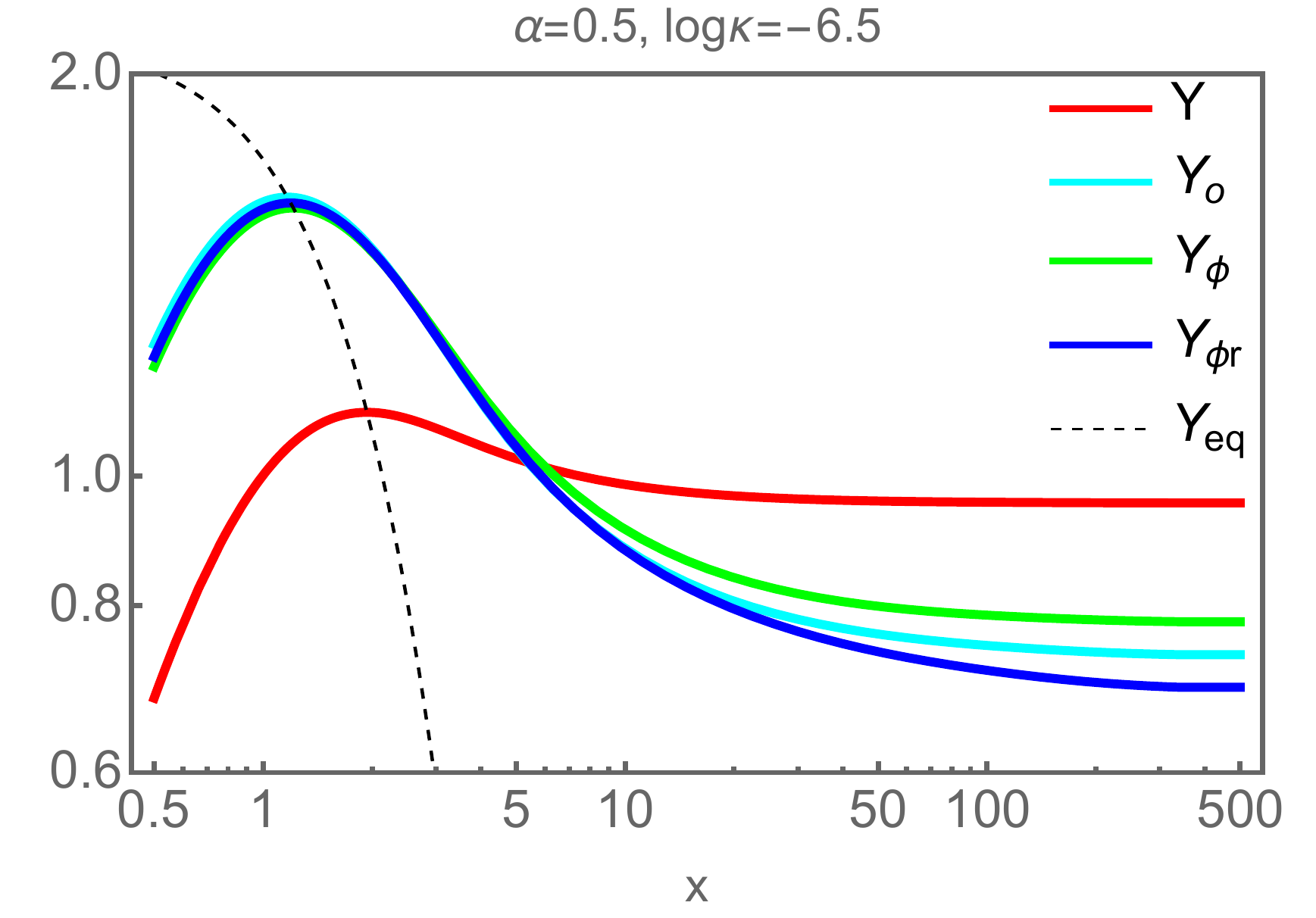}
    \end{minipage}}
    {\begin{minipage}{0.45\linewidth}
        \centering
        \includegraphics[width=1.0\linewidth]{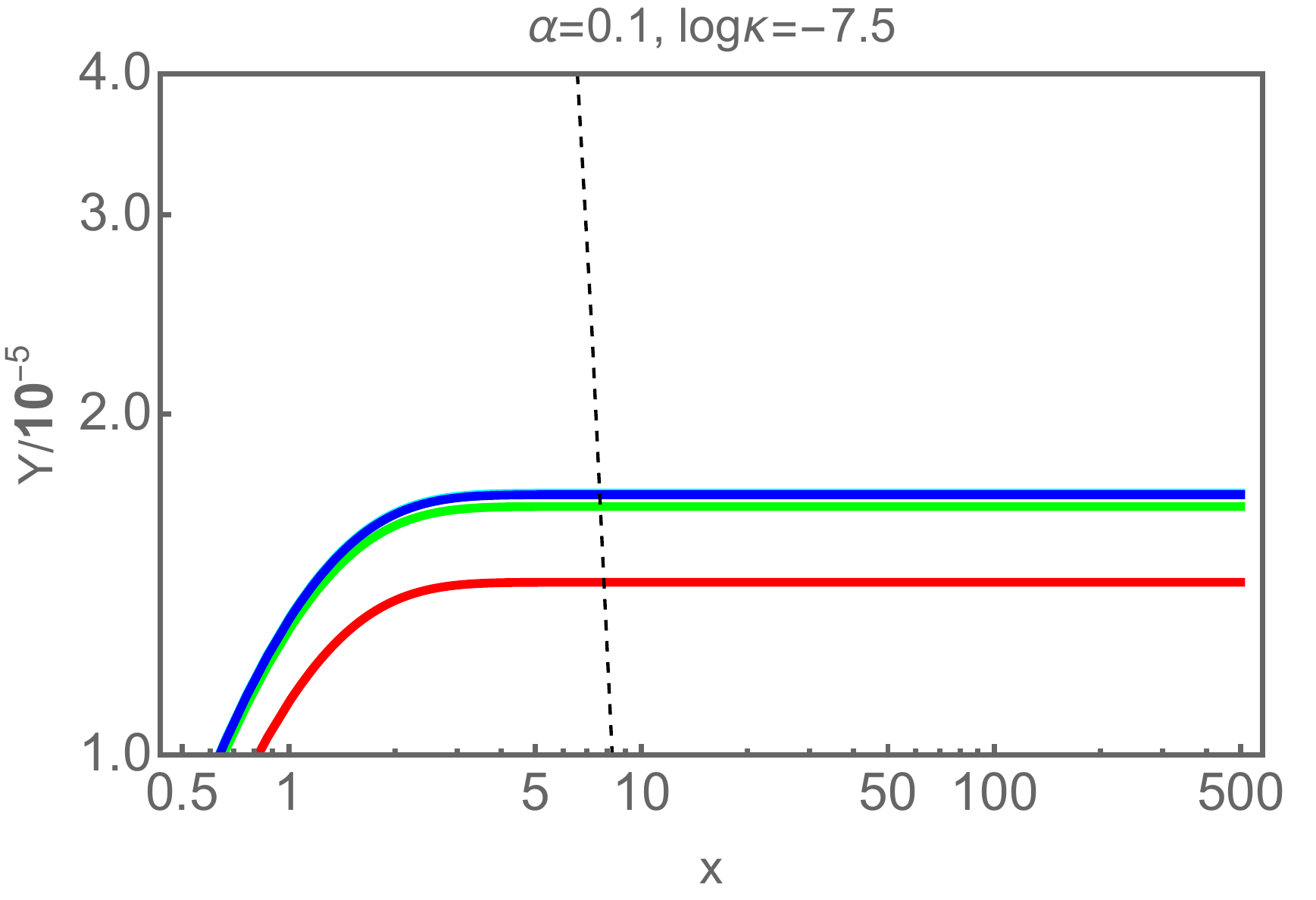}
    \end{minipage}}
    {\begin{minipage}{0.45\linewidth}
        \centering
        \includegraphics[width=1.0\linewidth]{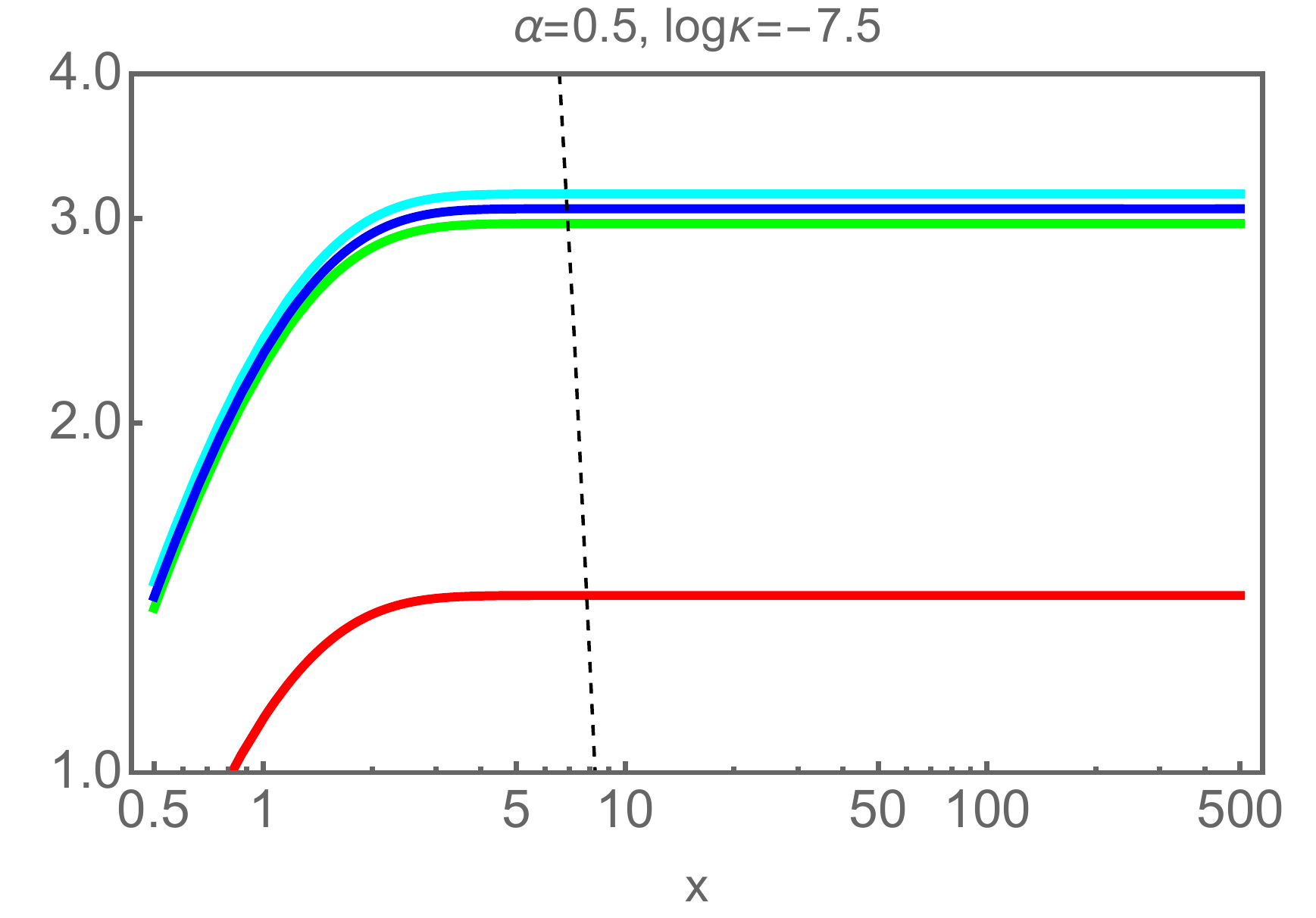}
    \end{minipage}}
    \caption{$Y, Y_o, Y_\phi, Y_{\phi r}, Y_{eq}$ are for no SE, massless mediator, $0.5m_\chi$ massive mediator, resonant mediator, and equilibrium value, respectively. These calculations are performed under the assumption of the $m_\chi = 1TeV,\ m_\phi=0.5 m_\chi$ (resonance $m_\phi=6\alpha m_\chi/\pi^2$ ). In the upper panels, large $\kappa$ enables the DM to reach thermal equilibrium and subsequently freeze out. Although SE leads to a maximum enhancement in the yield, it suppresses the final relic density. The bottom panels depict the case where $\kappa$ is small enough that the DM never reaches thermal equilibrium, yet the SE still significantly enhances the relic density. The SF used here is from Eq. \ref{eq:SF no mass} - \ref{eq:SF with mass}.} \label{fig: re freeze-out}
\end{figure}

\begin{figure}[t]
    \centering
    {\begin{minipage}{0.45\linewidth}
        \centering
        \includegraphics[width=1.0\linewidth]{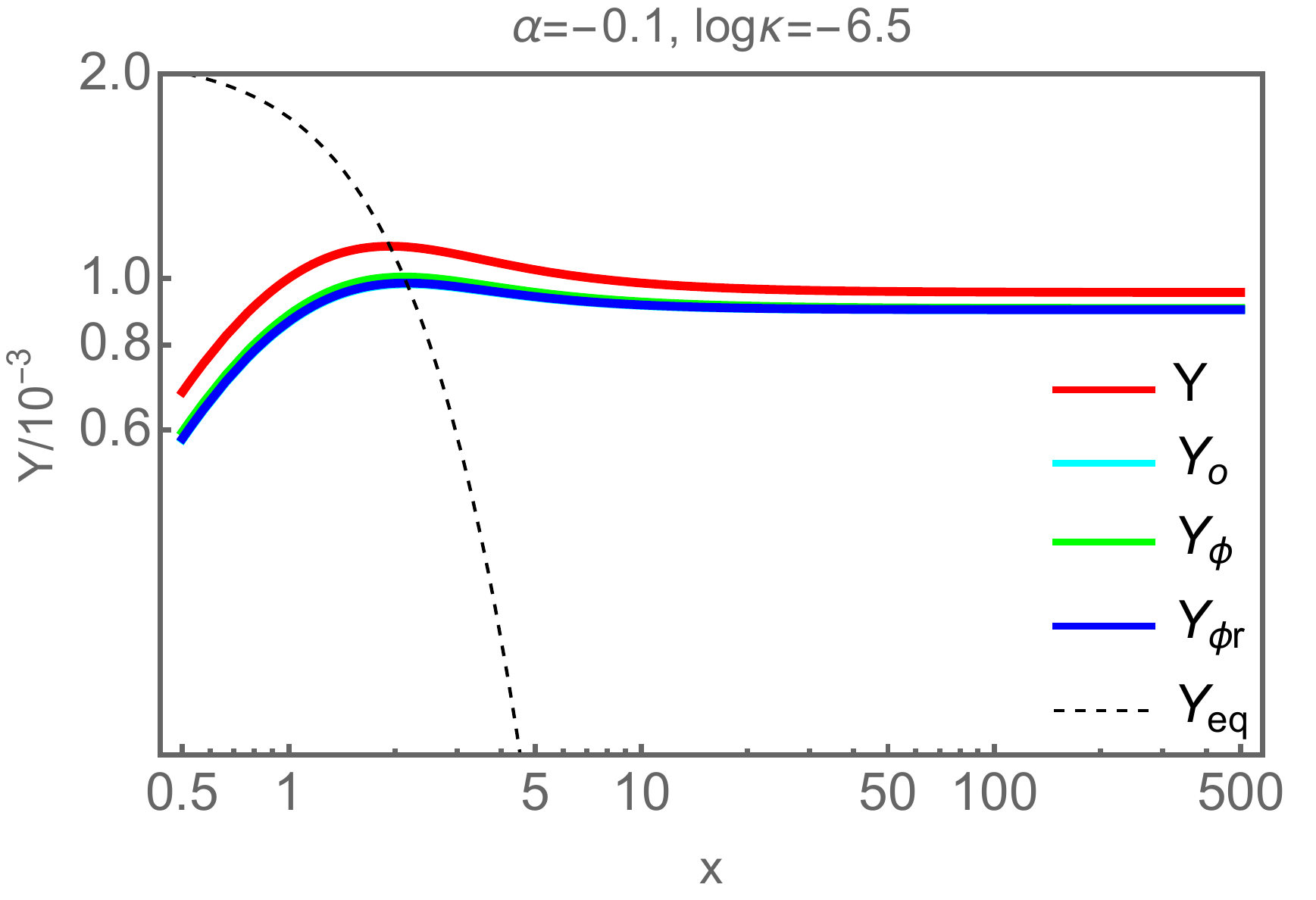}
    \end{minipage}}
    {\begin{minipage}{0.45\linewidth}
        \centering
        \includegraphics[width=1.0\linewidth]{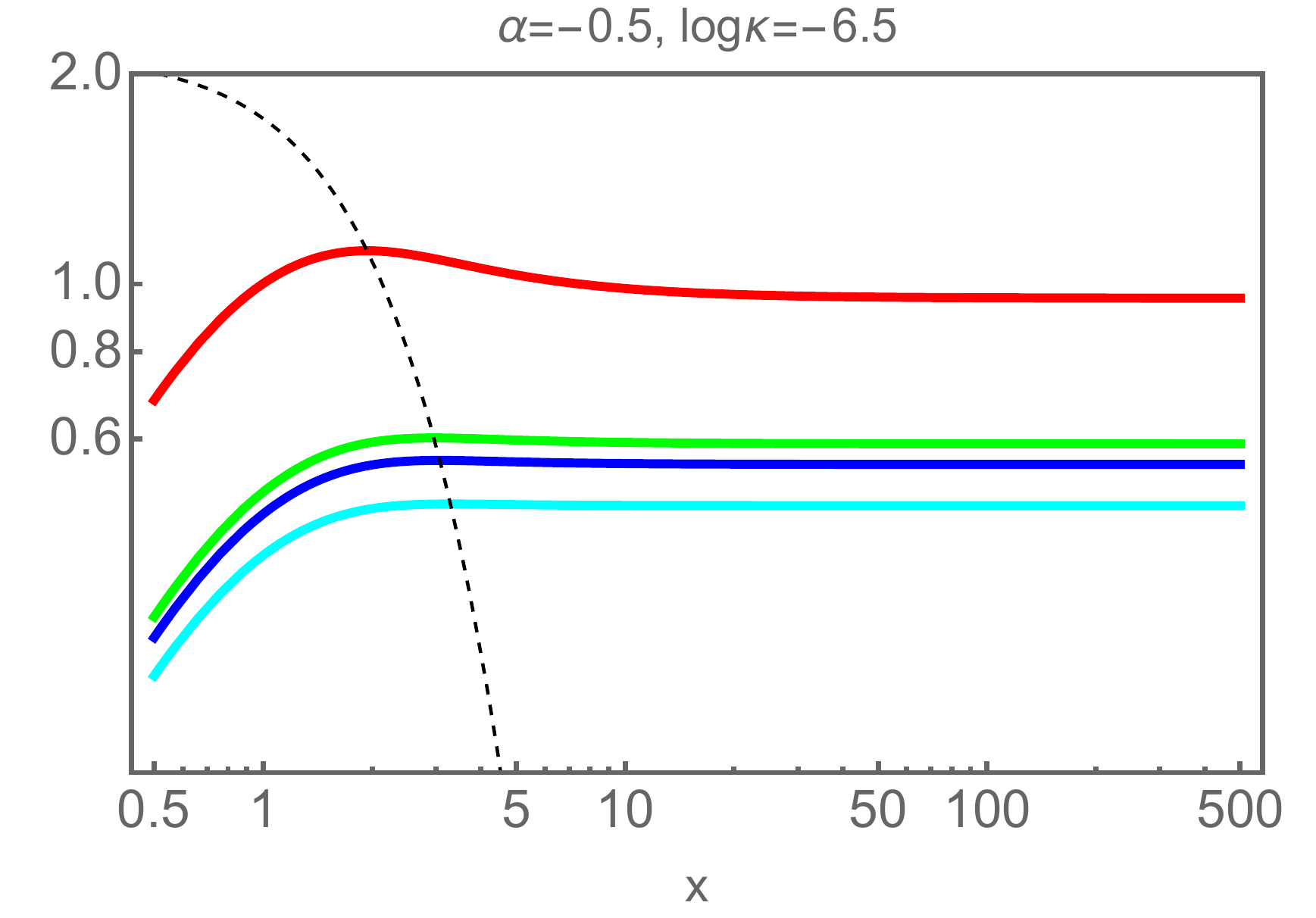}
    \end{minipage}}
    {\begin{minipage}{0.45\linewidth}
        \centering
        \includegraphics[width=1.0\linewidth]{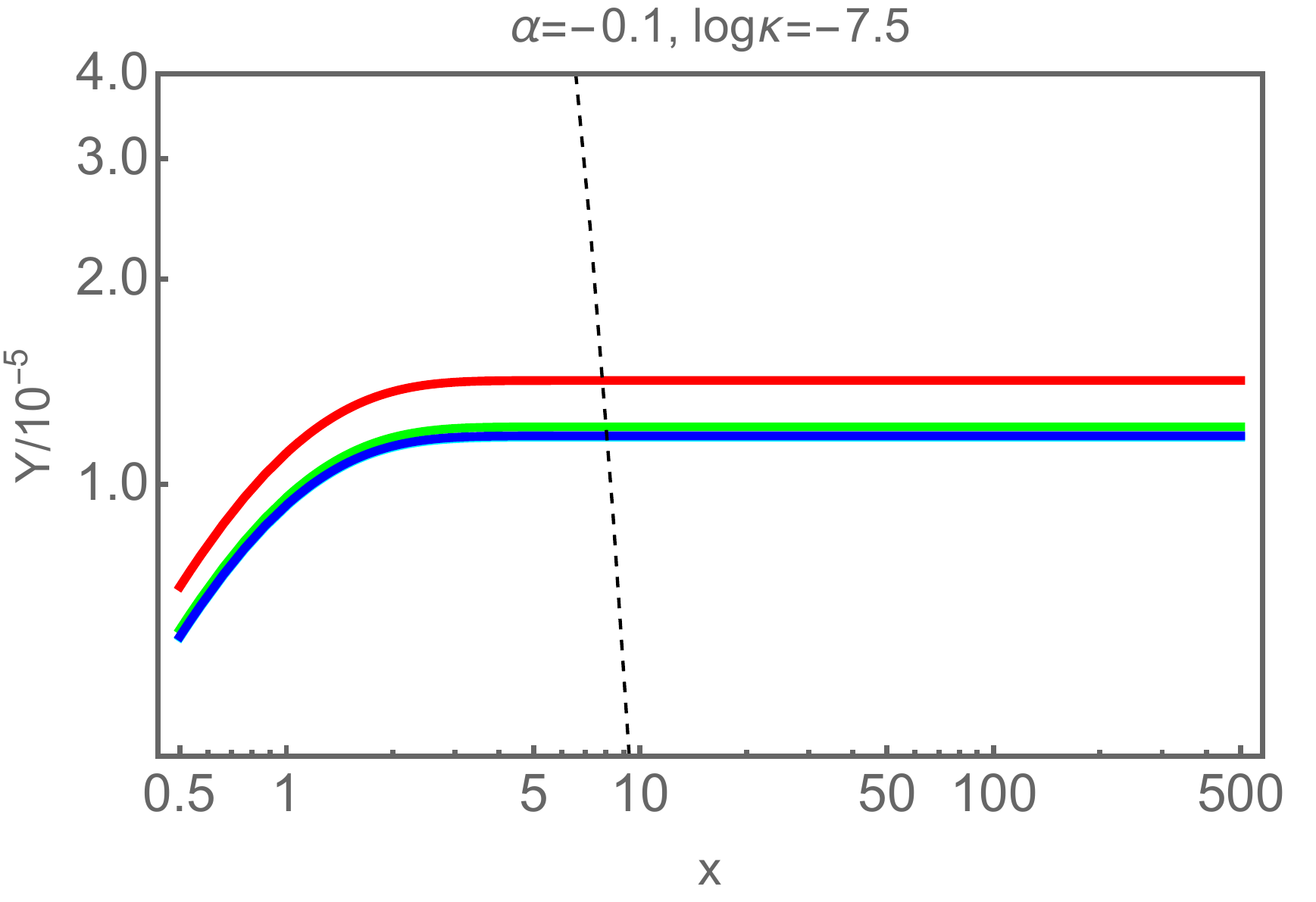}
    \end{minipage}}
    {\begin{minipage}{0.45\linewidth}
        \centering
        \includegraphics[width=1.0\linewidth]{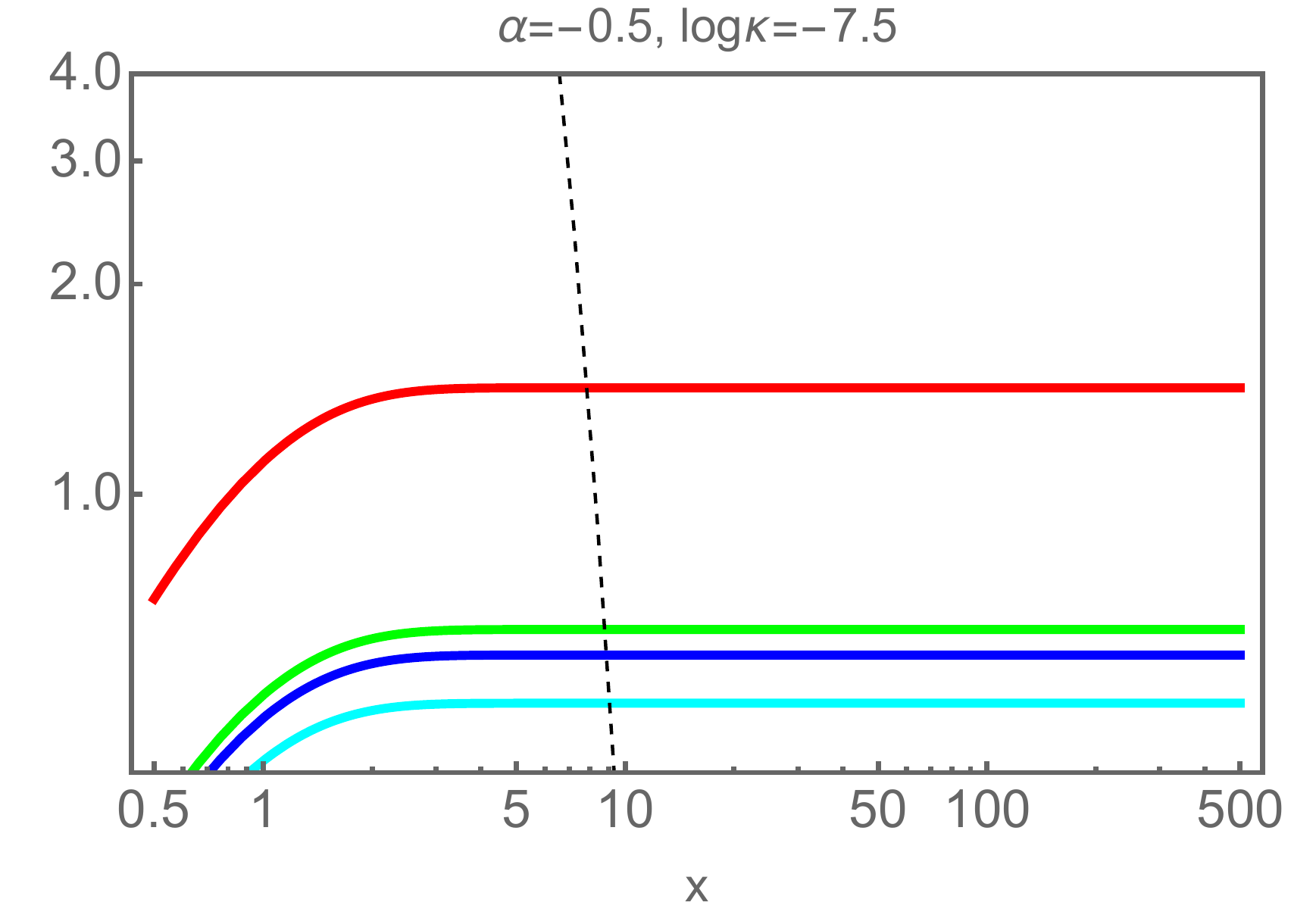}
    \end{minipage}}
    \caption{Same parameters as Fig \ref{fig: re freeze-out} but negative $\alpha$ for repulsive force. The SE shows a totally opposite influence compared to the attractive force. (Red line for reference). The repulsive force SE also shows an overall suppression.} \label{fig: re freeze-out2}
\end{figure}

Next, we will discuss the $\xi \to 1$ situation,  which indicates that the DM is in equilibrium with the VS. The critical temperature is determined by the condition:
\begin{align}
      \left \langle \sigma v _{\chi \to f } \right  \rangle n_\chi^{eq} = H(T),
\end{align}
where the Eq. \ref{eq:freeze-in condition2} changes its inequality sign. 
Ignoring the SE, and assuming $\left \langle \sigma v _{\chi \to f } \right  \rangle \sim \kappa^2/m_\chi^2$, we obtain the condition for the coupling constant between DM and VS [Appendix \ref{app: balance condition}]:
\begin{align}\label{eq: kappa value}
    \kappa^2 \sim \sqrt{\frac{4 \pi^3 g_*}{45}} \frac{x_f \ m_\chi} {m_{pl}}.
\end{align}
To demonstrate the impact of the SE on the DM abundance when condition \ref{eq:freeze-in condition2} is reversed, we set $z=1$, $\alpha=0.1$, and choose $\kappa$ to be approximately the value given in Eq. \ref{eq: kappa value}. We numerically solve the relic density Eq. \ref{eq: linear approx}, starting at $x=0.1$. The results are illustrated in Fig \ref{fig: re freeze-out} and Fig \ref{fig: re freeze-out2}.
If there are $2 \to n $ processes \cite{Heeba:2018wtf}, the Boltzmann equation is the same as Eq. \ref{eq: linear approx} when the detailed balance condition is applied. The previous conclusion still holds. The main difference is the signal of $\chi$ will be weaker, and the spectrum will be blurry \cite{Slatyer:2021qgc}.

In summary, the SE effect in the pure freeze-in region will increase the DM relic density if there is an attractive force, or decrease it if there is a repulsive force. The maximum modification of the relic density is likely to be no more than 20\% (Fig \ref{fig: SE for 1/s}) for TeV-scale DM while considering the constraint of the corrected relic DM density. As the portal coupling $\kappa$ increases, the influence of SE on the relic density will be opposite to that in the pure freeze-in region (upper panels of Fig \ref{fig: re freeze-out}). This means that the SE with an attractive force will decrease the relic density, while the SE with a repulsive force will increase the relic density. It is because the DM reaches its thermal equilibrium density that the enhancement of the cross-section will not further increase its density. However, it will increase the consumption of DM during the later freeze-out epoch.

\section{The SE in hidden dark sector}
\label{SE in hidden}
To make the SE correction significant, it requires a low-mass mediator to propagate the long-range force and a low enough temperature to reduce the relative velocity of the particles involved. If such a mediator exists in the VS or DS, and the sector is at a low enough temperature, it will result in a relatively large SE modification. Therefore, in the process of DM reannihilation or DS freeze-out, the correction of the cross-section by the SE will be larger than that in pure freeze-in. In this section, we will investigate the influence of the SE in the phases of DM reannihilation and DS freeze-out, considering the presence of a low-mass mediator in the DS.

The simplest DS consists of two kinds of hidden particles. The SM particles and DS particles form an intriguing triangular relationship \cite{Hambye:2019dwd}. Many models can be regarded as two-component DS models, for example, the dark photon and dark fermion model, dark $Z'$ model, and ALPs model \cite{Pospelov:2007mp, Fabbrichesi:2020wbt, Chu:2011be, Agashe:2014yua, Jaeckel:2014qea}... 
These two sectors might be in different thermal equilibriums, involving different thermodynamic quantities. We denote the temperature of the VS and the DS temperature as $T$ and $T'$, respectively. This is analogous to the difference in temperatures between the CMB and neutrino background, with $\xi = (4/11)^{1/3}$. Assuming that the DM particles $\chi$ and the mediator $\phi$ constitute the DS, and neglecting the variations in entropy and relativistic degrees of freedom, the general Boltzmann equation can be written as follows [see Appendix \ref{app: Bol with DS}]:
\begin{align}
    & \frac{H_{m_{\chi}}}{x s}  \frac{dY_\chi}{dx} = \left \langle \sigma v _{\chi \to  f } \right  \rangle \cdot (Y_\chi ^{eq \ 2} - Y_\chi^2) - \left \langle \sigma v _{\chi \to \phi } \right  \rangle \cdot  Y_\chi ^2 + \left \langle \sigma v _{ \phi \to \chi } \right  \rangle \cdot Y_{\phi}^2 \\
    & \frac{H_{m_{\chi}}}{x s}  \frac{dY_{\phi}}{dx} = \left \langle \sigma v _{\chi \to  \phi } \right  \rangle \cdot  Y_\chi ^2 - \left \langle \sigma v _{ \phi \to \chi } \right  \rangle \cdot Y_{\phi}^2.
\end{align}

\subsection{Thermalization of the dark sector with SE}
\begin{figure}[t]
    \centering
    {\begin{minipage}{0.45\linewidth}
        \centering
        \includegraphics[width=1.0\linewidth]{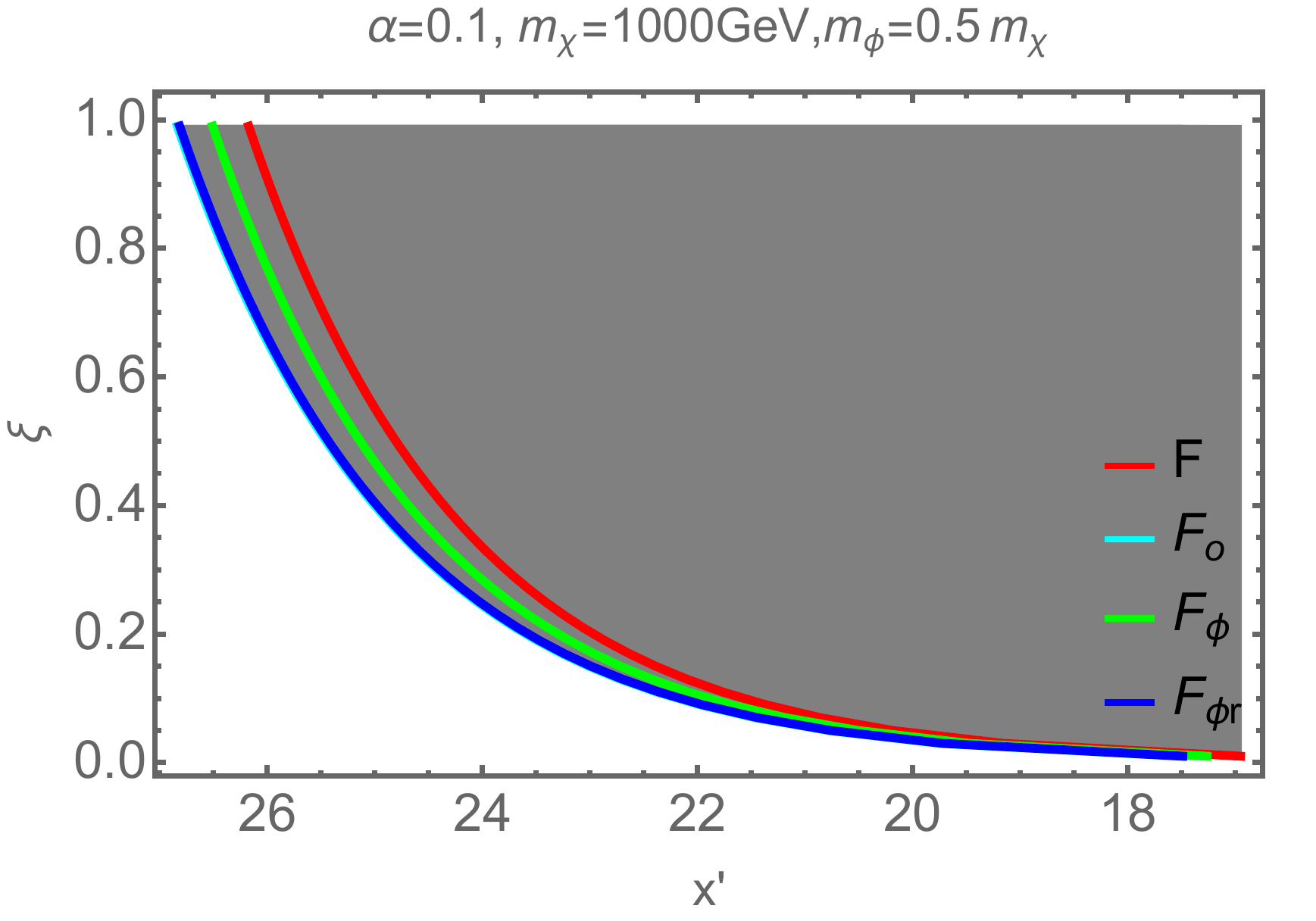}
    \end{minipage}}
    {\begin{minipage}{0.45\linewidth}
        \centering
        \includegraphics[width=1.0\linewidth]{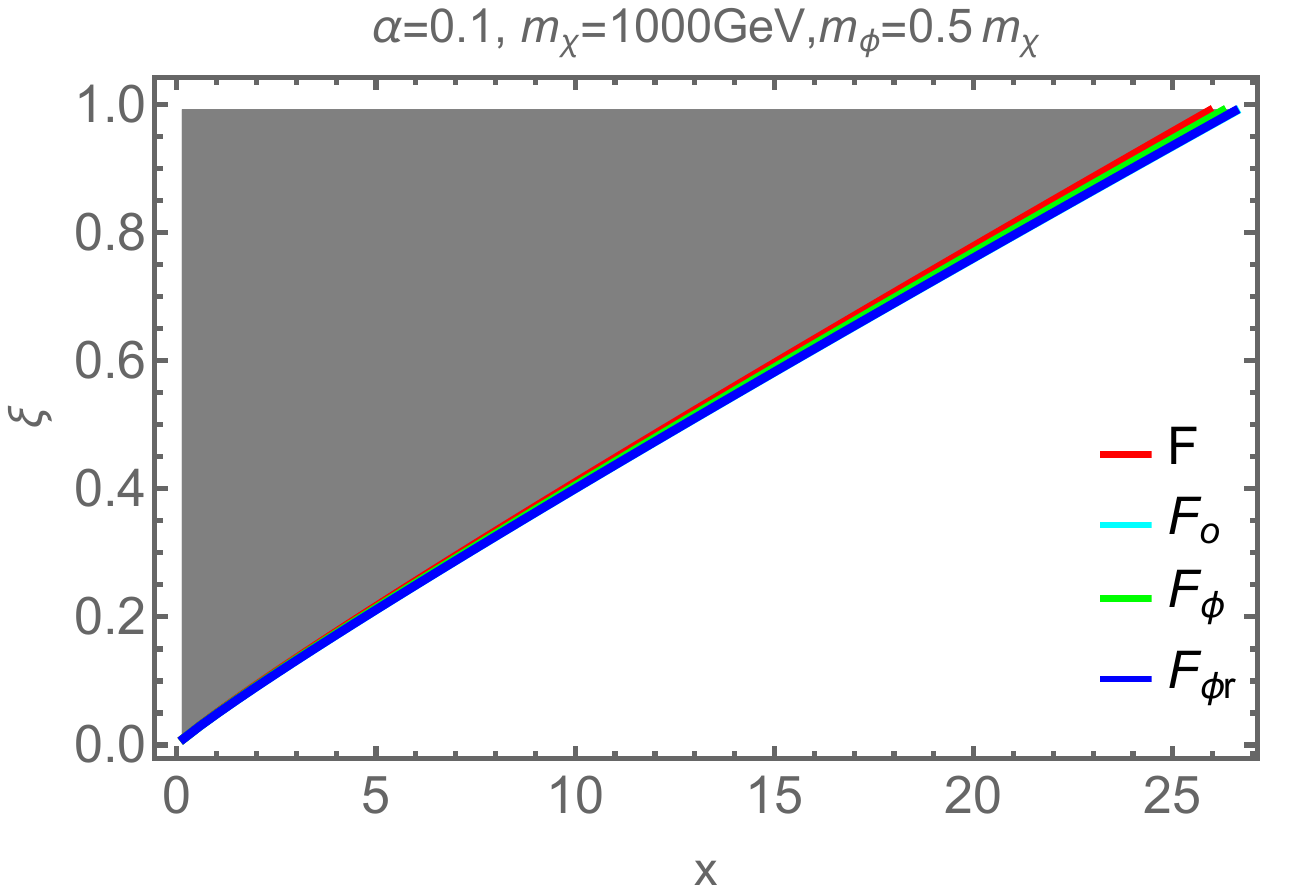}
    \end{minipage}}
    \caption{$F, F_o, F_\phi, F_{\phi r}$ represent the cases without SE, with a massless mediator, with a mediator of $0.5m_\chi$ mass, and with a resonant mediator, respectively. 
    In the left panel, the curves correspond to the zero points of Eq \ref{eq: zero of Fi}.
    The shaded area represents the DS temperature range where $\phi$ equilibrium is established. Taking the attractive force SE into consideration results in faster attainment of DS thermal equilibrium for $\phi$. The right panel (Eq \ref{eq: x vs xi}) displays the VS temperature (shaded) when the $\phi$ is in the DS equilibrium. The curves $F_o$ and $F_{\phi r}$ overlap. The relative positions of the curves are switched when the repulsive force SE is imposed.} \label{fig: DS thermal equilibrium}
\end{figure}

Before the time of the decoupling, the DS is gradually thermalized by the VS. The energy is transferred to the DS from the VS through the connector/portal reaction $f  \longleftrightarrow  \chi$. When the reaction rate of $\phi \to \chi$ catches up with the universe's expansion rate, 
\begin{align}
    \left \langle \sigma v _{  \phi \to \chi}  \right  \rangle  n_\phi > H(T),
\end{align}
it means that $\phi$ is in the equilibrium of DS. Otherwise, the reaction with the $\phi$ particles can be neglected, and we only need to consider the evolution of $\chi$ without including the $\phi$. In this case, it reverts to the result of Sec \ref{sec: pure freeze-in}. 
To keep the $\phi$ in the DS equilibrium, the reaction rate should be larger than the Hubble rate.
With the assumption of the DS with equilibrium temperature $T'$, the condition is as follows: 
\begin{align}
     [\left \langle \sigma v _{\chi \to  \phi }  \right  \rangle n^{eq}_\chi ]_{T'} > H(T).
\end{align}
Considering the SE of $\chi$ annihilation to $\phi$, which contributes to the $\sim v^{-1}$ or $v^{-2}$ correction on cross-section at the low DS temperature, the LHS of the equation will be larger. As a result, the DS reaches thermal equilibrium faster. We define a function of $x'$ as follows:
\begin{align}\label{eq: zero of Fi}
    F_i(x') = ln[ \frac{\xi^2 x'^2}{ H_{m_\chi}} [\left \langle \sigma v _{\chi \to  \phi }  \right  \rangle n^{eq}_\chi ]_{T'} ].
\end{align}
Here, $F_i(x') > 0$ is the condition for the $\phi$ to be in thermal equilibrium within the DS. The corresponding time of equilibrium established is given by: 
\begin{align}\label{eq: x vs xi}
    x = \xi \ x'|_{F_i(x')=0}.
\end{align}
Assuming the cross-section for the $\chi$ annihilation to $\phi$ is given by:
\begin{align}
    \sigma_{\chi \to  \phi } = \frac{\alpha^2}{4m^2_\chi \tilde{s}},
\end{align}
Fig \ref{fig: DS thermal equilibrium} shows the possible $\xi$  (shade) when the DS thermal equilibrium established. 

\begin{figure}[t]
    \centering
    {\begin{minipage}{0.45\linewidth}
        \centering
        \includegraphics[width=1.0\linewidth]{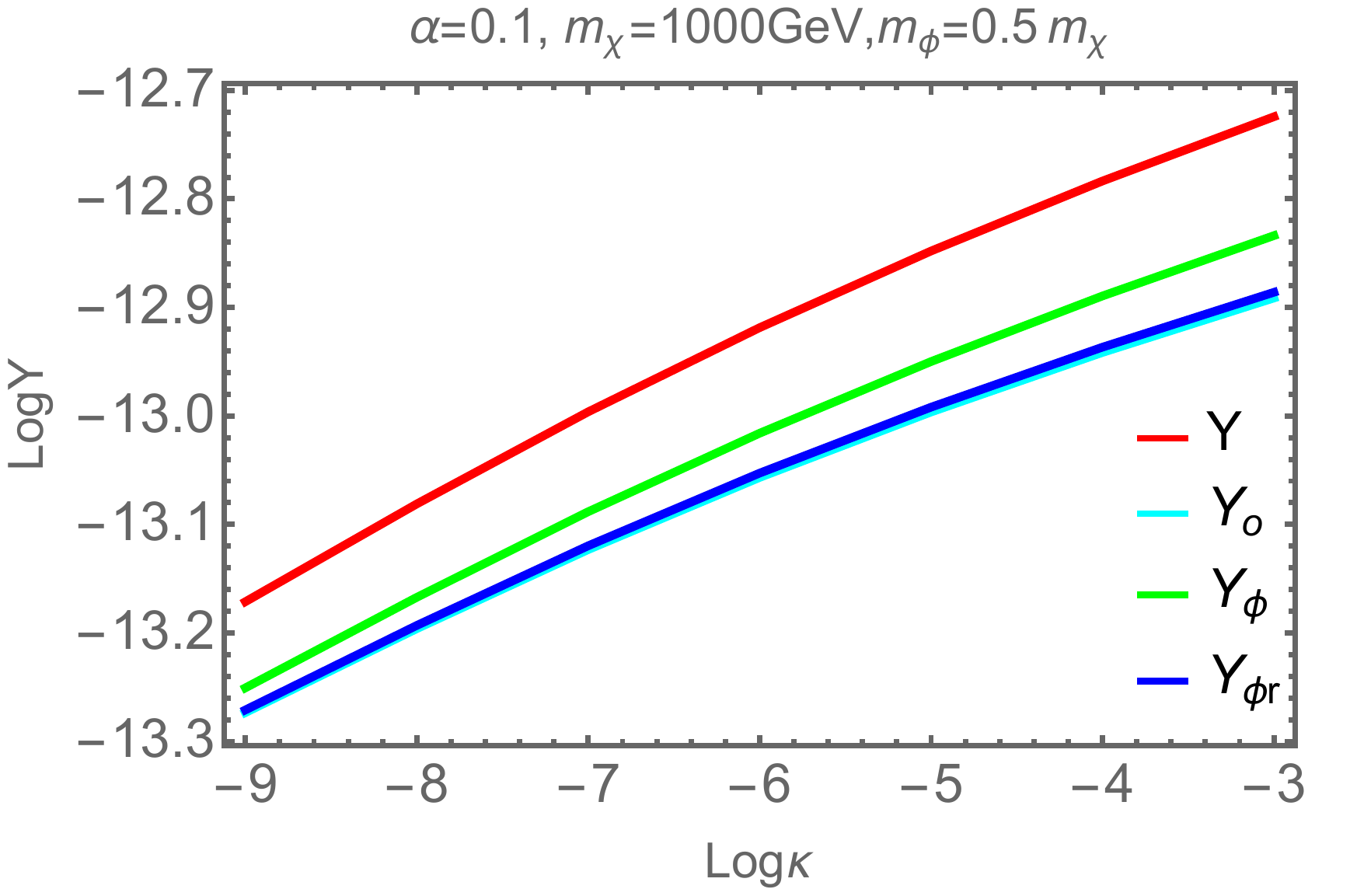}
    \end{minipage}}
    {\begin{minipage}{0.45\linewidth}
        \centering
        \includegraphics[width=1.0\linewidth]{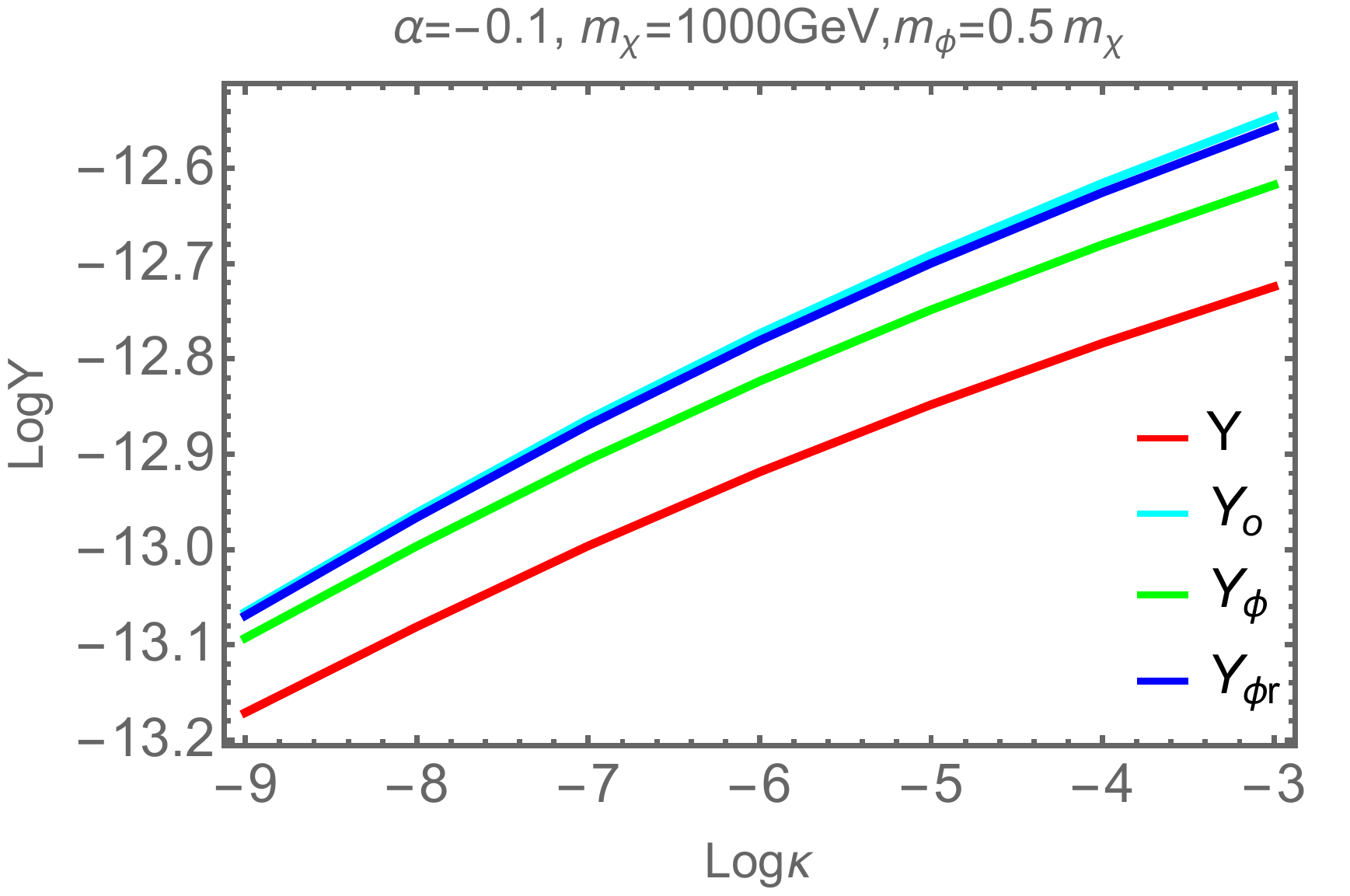}
    \end{minipage}}
    \caption{$Y, Y_o, Y_\phi, Y_{\phi r}$ stand for the cases without SE, with a massless mediator, with a mediator of $0.5m_\chi$ mass, and with a resonant mediator, respectively. The SE has an overall suppression/enhancement on the final DM abundance. The changes in $Y$ are small as $\kappa$ varies over a large range. This region corresponds to the ``mesa'' in the phase diagram.}\label{fig: Y_including_DS}
\end{figure}

If $\phi$ couples strongly enough to $\chi$ and is in the DS thermal bath, the Boltzmann equation is reshaped as follows [Appendix \ref{app: Bol with DS}] :
\begin{align}
    & \frac{ H_{m_{\chi}}}{xs} \frac{dY_\chi}{dx} = \left \langle \sigma v _{\chi \to  f } \right  \rangle \cdot Y_\chi ^{eq \ 2}  + {\left \langle \sigma v _{\chi \to  \phi } \right  \rangle} \cdot  ( \xi^6 \ Y_\chi^{eq \ 2} - Y_\chi ^2).
\end{align}
Fig \ref{fig: DS thermal equilibrium} displays the region where $\phi$ is in the DS thermal equilibrium and the corresponding possible values of $\xi$ at different $x$. The $\xi^6$ suppression term can be disregarded when $\xi$ is sufficiently small, such as at $x<10$. Subsequently, the Boltzmann equation becomes:
\begin{align}
    & \frac{ H_{m_{\chi}}}{xs} \frac{dY_\chi}{dx} = \left \langle \sigma v _{\chi \to  f } \right  \rangle \cdot Y_\chi ^{eq \ 2}  - \left \langle \sigma v _{\chi \to  \phi } \right  \rangle \cdot   Y_\chi ^2.
\end{align}
This equation can be solved by introducing an effective cross-section \cite{Chu:2011be}:
\begin{align}
    \left \langle \sigma v _{eff} \right  \rangle_i = \sqrt{\left \langle \sigma v _{\chi \to  f } \right  \rangle_i \left \langle \sigma v _{\chi \to  \phi } \right  \rangle_i },
\end{align}
then the final DM relic density can be evaluated as follows:
\begin{align}
    & x_f= \log B -\frac{1}{2} \log (\log B)\\
    & B =  0.038 \frac{g_\chi}{\sqrt{g_{*eff}}} m_{p l} m_\chi \left\langle\sigma v _{e f f}\right\rangle_i c(c+2)
\end{align}
\begin{align} \label{eq: reann Y}
    & Y_\chi(x_f)  = \frac{3.79 x_f \sqrt{g_{*eff}}}{g_{\star s} m_{p l} m_\chi \left \langle \sigma v _{\chi \to  \phi } \right  \rangle}_i \\
    & \Omega_{\chi} h^2  = \frac{1.07 \times 10^9 \ x_f \sqrt{g_{*eff}} \cdot \mathrm{GeV}^{-1}}{g_{* s}  m_{pl} \left\langle\sigma v _{\chi \to  \phi} \right\rangle_i }
\end{align}
where $g_{*eff}$ is the number of relativistic degrees of freedom including the DS [Appendix \ref{app: Bol with DS}]; $g_\chi$ is the degrees of freedom of $\chi$; $c$ is a constant typically chosen as $c(c + 2) = 1$. 
Taking $g_\chi = 1$ and $g_{*eff} = 108.75$, the calculated $yield$ under the influence of SE is shown in Fig \ref{fig: Y_including_DS}.
The SE will suppress/enhance the final relic abundance of DM, while the change of the value of $\kappa$ does not affect the value of $Y_\chi$ too much. This behavior is visually represented by a ``mesa'' shape in the phase diagram.

In summary, in the reannihilation region, DM is thermalized, but DM always maintains its equilibrium with other species particles of DS (for example mediator $\phi$) due to the relatively large internal/portal reaction rate.
The DS bath does not require the same temperature as the VS, but its minimal temperature depends on the portal coupling $\kappa$, which decides how much energy is transferred from the VS to the DS. The increase of $\kappa$  will not significantly influence the DM relic density, because the DM particle annihilation to its mediator keeps the number density stable. The SE will enhance/suppress the cross-section of annihilation, causing the number density of DM to decrease/increase (Fig \ref{fig: Y_including_DS}).

\subsection{Freeze-out of dark sector}
\label{sec: Freeze-out of dark sector}

\begin{figure}[t]
    \centering
    {\begin{minipage}{0.45\linewidth}
        \centering
        \includegraphics[width=1.0\linewidth]{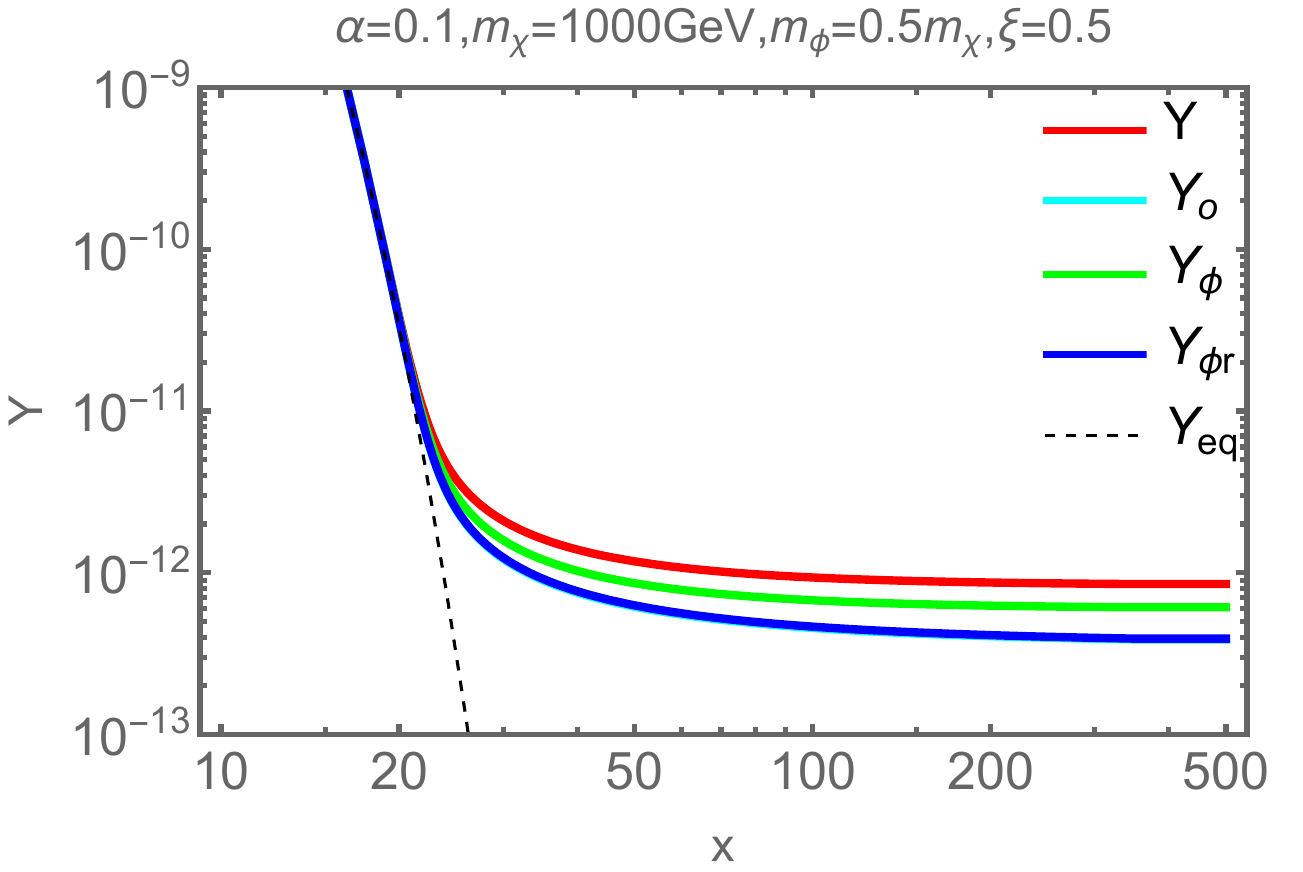}
    \end{minipage}}
    {\begin{minipage}{0.45\linewidth}
        \centering
        \includegraphics[width=1.0\linewidth]{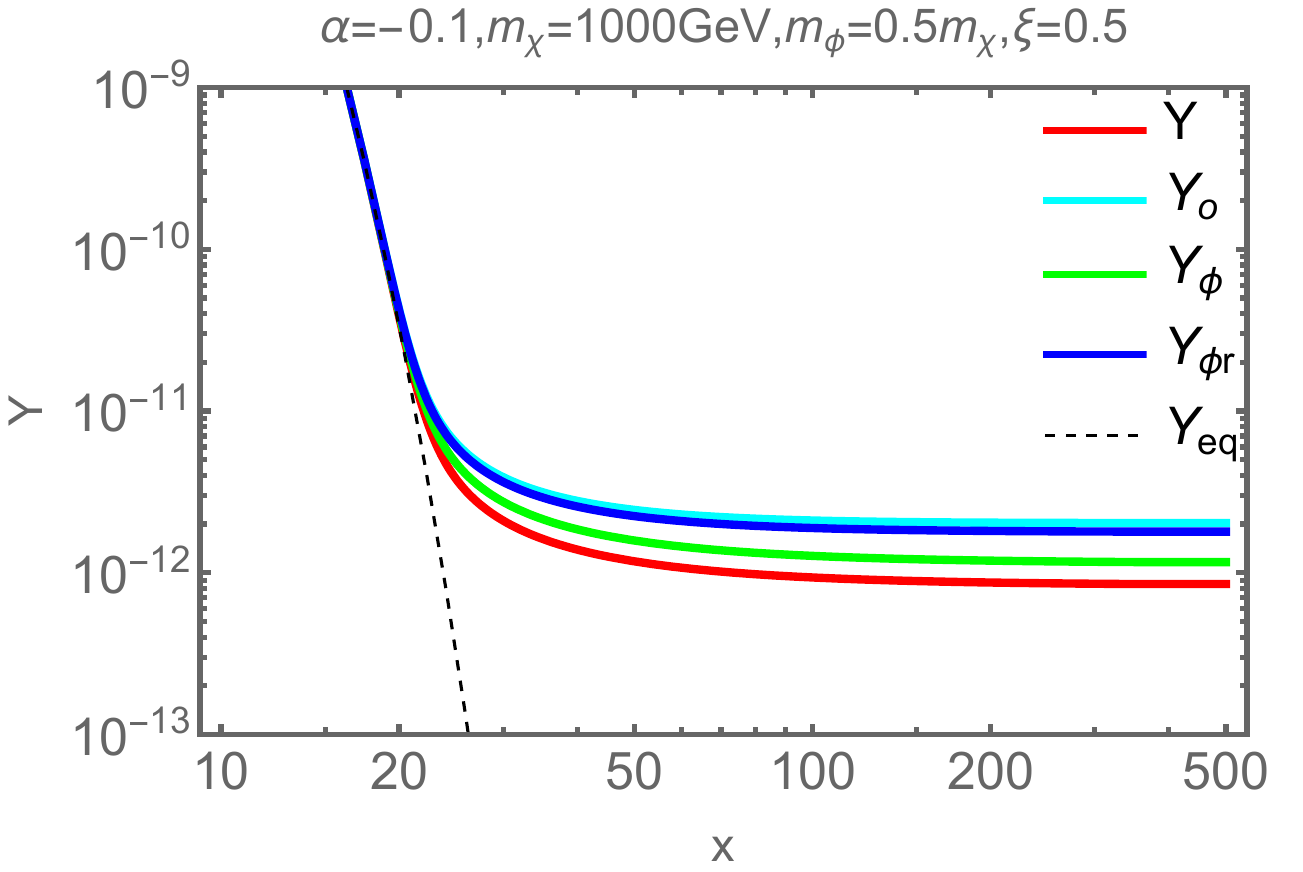}
    \end{minipage}}
    \caption{The examples of the DS freeze-out with $\xi=0.5$. $Y, Y_o, Y_\phi, Y_{\phi r}, Y_{eq}$ represent the relic abundances for the cases without the SE, with a massless mediator, with a mediator of $0.5m_\chi$ mass, with a resonant mediator, and the equilibrium value, respectively. It shows the DS freeze-out earlier than usual freeze-out. Noted that the $Y_o$ and $Y_{\phi r}$ curves overlap in the left panel.}\label{fig: diff xi}
\end{figure}

The initial DM density is assumed to be negligible $n_\chi \sim 0$,  so there is a process of thermalization in the DS. If the DS bath reaches and remains at its equilibrium after freeze-in, the final DM relic density is decided by the DS freeze-out. One thing that needs to be emphasized is the value of $\xi$ at the end of freeze-in because it provides the initial condition for the subsequent DS freeze-out. Other factors remain the same as in the usual freeze-out scenario.  The governing Boltzmann equation is 
\begin{align}
    & \frac{ H_{m_{\chi}}}{xs} \frac{dY_\chi}{dx} =  \left \langle \sigma v _{\chi \to  \phi } \right  \rangle'_i \cdot  ({Y'}_\chi^ {eq \ 2} - Y_\chi ^2),
\end{align}
where the $'$ denotes the temperature $T'$. After decoupling between the VS and the DS, the temperature ratio $\xi$ remains constant before the DS freeze-out, as we assume. The initial density of $\chi$ is 
\begin{align}
    {Y'}_\chi^ {eq} = \xi ^3 {Y}_\chi^ {eq}.
\end{align}
From Fig \ref{fig: DS thermal equilibrium}, we know $\xi$ should be larger than $\sim 0.5$ at the typical freeze-out time $x=20$. The $\xi^3$ suppression of ${Y'}_\chi^{eq}$ indicates that $\xi$ should not be too small if the DS freeze-out occurred. Fig \ref{fig: diff xi} shows the impact of the SE on the DS freeze-out.
The corresponding freeze-out time can also be calculated by \cite{Chu:2011be}
\begin{align} \label{eq: dark freeze-out Y}
        & x_f= \xi \ln B - \frac{1}{2} \xi \ln (\xi \ln B) \\
        & B =  0.038 \xi^{5/2} \frac{g_\chi}{\sqrt{g_{*eff}}} m_{p l} m_\chi \left\langle\sigma v_{\chi \to \phi}\right\rangle_i c(c+2).
\end{align}

The only difference in the DS freeze-out compared to the usual freeze-out is that the initial number density of DM (or the temperature of the DS bath) is not equal to that of the VS. Usually, the temperature of the DS bath is not higher than that of the VS bath, which is the main reason for the early freeze-out of the DS compared to the usual freeze-out. The SE in this region will not be too much different from the region of usual freeze-out, that is the enhancement of cross-section will increase the DM consumption, and vice versa (Fig \ref{fig: diff xi}).

\section{The Model}
\label{sec: Model}

In the previous sections, we analyzed the SE in the different freeze-in  ``phases'' and roughly calculated the related relic density. In this section, we will concretely calculate the corrected cross-section in a suitable model and illustrate the modification of the final DM abundance. 
We consider the SE correction in a $2 \to 2$ reaction. The magnitude of the SE correction is related to the structure of the cross-section, for instance, 4-point interactions, the s-channel \cite{Backovic:2015soa, Englert:2016joy}, and t-channel \cite{Becker:2022iso, Arina:2020tuw, Arina:2020udz} DM. We will calculate the relic density of two specific models in the following.

\subsection{U(1) dark photon}

In this section, we will calculate the SE correction on the relic density of the U(1) dark photon model \cite{Holdom1986TwoUA}. In this model, a light dark photon is coupled with the DM particle $\chi$, and the interaction part of the Lagrangian is given by:
\begin{align}
    \mathcal{L}_{int} = - \frac{\kappa}{2} F^\prime _{\mu \nu} F^{\mu \nu} + \sqrt{4 \pi \alpha'} \ \bar{\chi} A \!\!\!/^\prime \chi.
\end{align}
The DM production channel is
\begin{align}
    f \ \bar{f} \to \gamma \to \gamma' \to \chi \ \bar{\chi}.
\end{align}
In the DS, the $\chi$ pair is consumed by the process
\begin{align}
    \chi \ \bar{\chi} \to 2 \ \gamma^\prime.
\end{align}
\begin{figure}[t]
        \centering
    {\begin{minipage}{0.3\linewidth}
            \centering
            \includegraphics[width=1.0\linewidth]{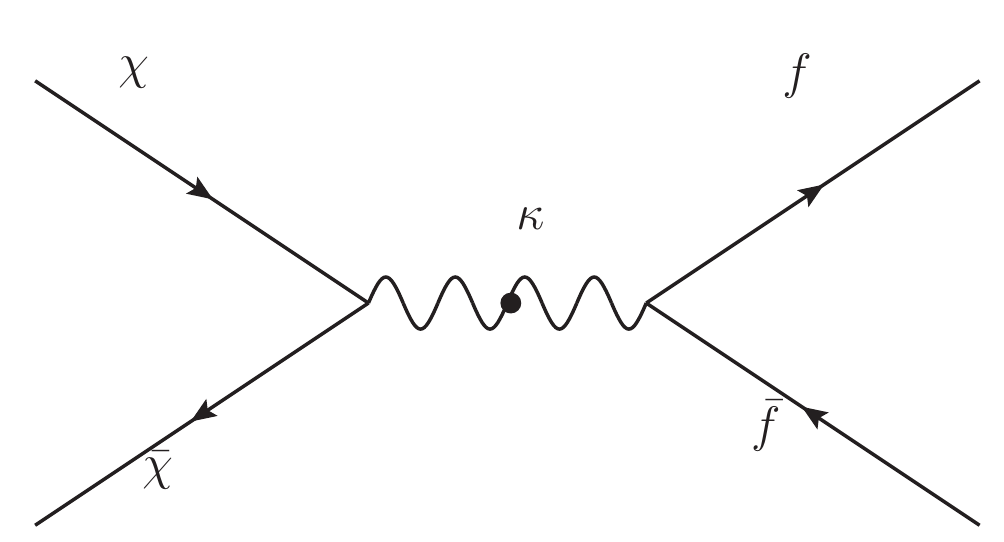}
    \end{minipage}}
    \hspace{0in}
    {\begin{minipage}{0.3\linewidth}
        \centering
        \includegraphics[width=1.0\linewidth]{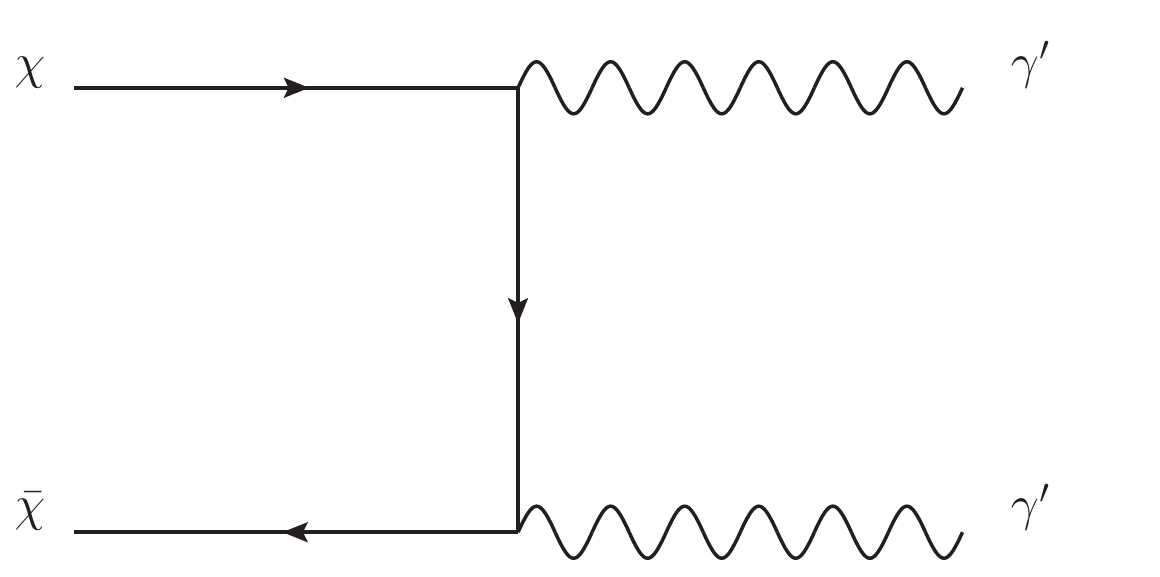}
    \end{minipage}}
    \caption{The left panel is the portal between the DS and the VS. The right panel represents the self-interaction within the DS, including the $t$ and $u$ channels. The corresponding tree-level diagram will be depicted in the middle circle of Fig \ref{fig: SM enh}.}\label{fig: U1 dark photon}
\end{figure}
The corresponding reactions are shown in Fig \ref{fig: U1 dark photon}, and the cross-sections (calculated by feyncalc \cite{Shtabovenko:2020gxv}) are as follows:
\begin{align}
    \sigma_{\chi \to f} & = \frac{\pi \ \alpha \ \alpha_{em} \ \kappa^2 }{12 m_\chi ^2 } \frac{(2\tilde{s}+1)(2\tilde{s}+z)}{\tilde{s}^3} \frac{\sqrt{1-z \tilde{s}^{-1}}}{\sqrt{1-\tilde{s}^{-1}}} \to \frac{\alpha \kappa^2 (\tilde s + 1/2)^2}{4 m_\chi^2 \tilde s^3} \\
    \sigma_{\chi \to \gamma'} & \approx \frac{ \pi \alpha^2}{2 m^2_{\chi} \tilde{s}}\frac{\sqrt{1- (m_\phi/m_\chi)^2\tilde{s}^{-1}}}{ \sqrt{1-\tilde{s}^{-1}}} \to \frac{\alpha^2}{4 m_\chi^2 \tilde s},
\end{align}
where $z \equiv m_f^2/m_\chi^2$; the last arrows stand for the constants are absorbed into $\kappa$ or $\alpha$, and $z$ was taken equal 1. 

The purpose of rescaling the coupling constants is to make the parameter spaces of both models similar and comparable. This rescaling also simplifies the cross-section and makes it consistent with the previous section's discussion. The $\alpha_{em}$ or $\alpha_{weak}$ coupling can provide the SE correction in the $\sigma_{\chi \to f}$ channel. The final result is similar to the FIMP model discussed below, except for in the freeze-in region where the dominant cross-section $\sigma_{\chi \to f}$ is not only determined by $\kappa$ but also $\alpha$. The SE correction on relic density is shown in Fig \ref{fig: correct omega}.


\subsection{$2\rightarrow2$ point contact FIMP }
\begin{figure}[t]
        \centering
    {\begin{minipage}{0.3\linewidth}
            \centering
        \includegraphics[width=0.9\linewidth]{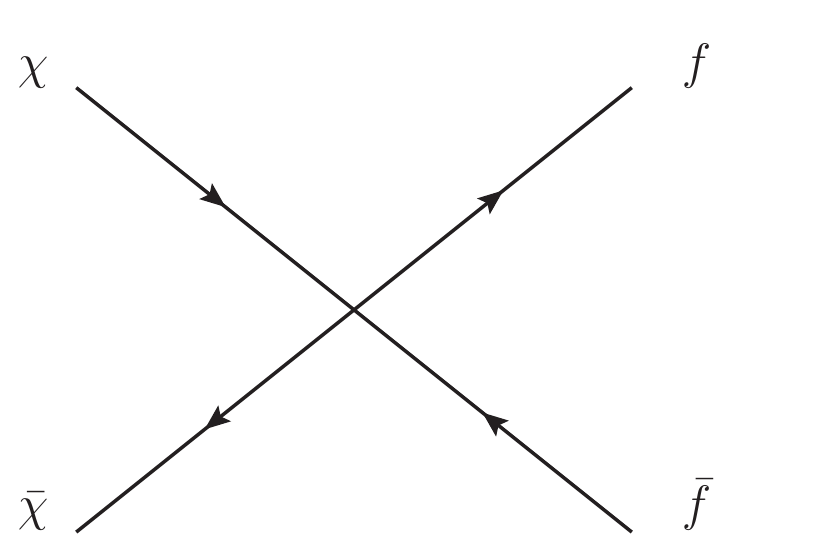}
    \end{minipage}}
    {\begin{minipage}{0.3\linewidth}
        \centering
        \includegraphics[width=0.9\linewidth]{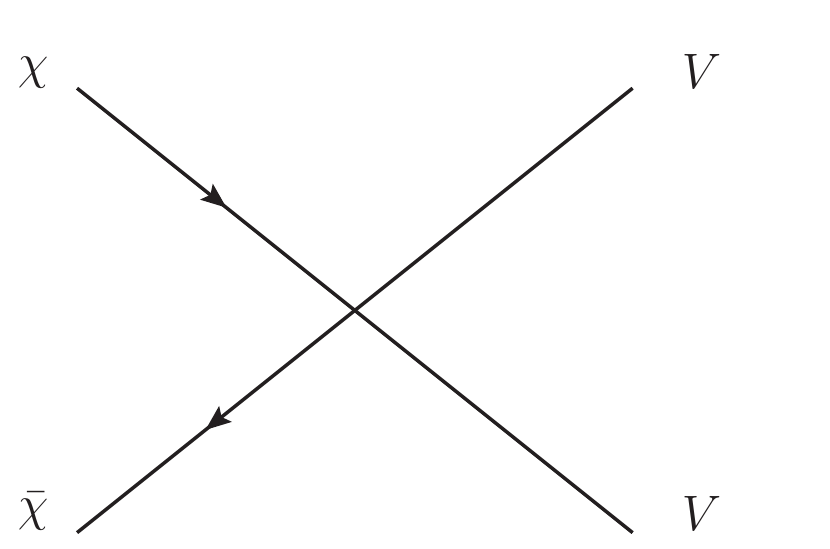}
    \end{minipage}}
    \caption{The left panel represents the portal between the DS and the VS. The right panel is DS self-interaction. There are 3 $V$ producing channels, but $t$ and $u$ channels are DM mass suppressed, so we only consider the 4-point contact channel. The corresponding tree-level diagram will be put in the middle circle of Fig \ref{fig: SM enh}.}\label{fig: 2to2 point contact FIMP}
\end{figure}

In the FIMP model, $\chi$ is a scalar that quartically interacts with the bath particle $f$. There is a scalar QED-like coupling between $\chi$ and the real vector boson $V$. The interaction part of the Lagrangian is summarized as follows:
\begin{equation}
    \mathcal{L}_{int}=|D_\mu \chi|^2 + \frac{1}{2}  \kappa \ |\chi|^2  \ |f|^2,
\end{equation} 
where $D_\mu=\partial_\mu - i \sqrt{4 \pi \alpha} V_\mu$ is the covariant derivative. The coupling constant of the four-point interaction is denoted by $\kappa$. The symbol $V_\mu$ represents the real vector boson. The DM particle $\chi$ is produced via an annihilation process $f \ \bar f \rightarrow \chi \ \bar \chi$ and consumed by the process $\chi \ \bar \chi \rightarrow 2 \ V$, as illustrated in Fig \ref{fig: 2to2 point contact FIMP}.
The corresponding cross-sections are given by \cite{Wang:2022avs}:
\begin{align}
    \sigma_{\chi \to f} & = \frac{\kappa^2}{32 m_\chi^2 \pi \tilde{s}}\frac{\sqrt{1-z \tilde{s}^{-1}}}{\sqrt{1-\tilde{s}^{-1}}} \to \frac{\kappa^2}{4m_\chi^2 s}\\
    \sigma_{\chi \to V} & \approx \frac{2 \pi \alpha^2}{m_\chi^2 \tilde{s}}\frac{\sqrt{1- (m_\phi/m_\chi)^2\tilde{s}^{-1}}}{ \sqrt{1-\tilde{s}^{-1}}} \to \frac{\alpha^2}{4m_\chi^2 s},
\end{align}
where the last arrows indicate that the constants have been absorbed into the parameters $\kappa$ or $\alpha$, and $z$ is taken to be equal to 1.

According to Eq. \ref{eq: simplest Y}, \ref{eq: reann Y}, \ref{eq: dark freeze-out Y}, and \ref{eq: dark freeze-out Y} with $\xi = 1$ and the replacement of $\left\langle\sigma v_{\chi \to \phi}\right\rangle_i$ to $\left\langle\sigma v_{\chi \to f}\right\rangle_i$, the relic density on four different phases can be calculated. The ``phase transition'' conditions, similar to those in \cite{Chu:2011be}, are given below:
\begin{align} 
    I \to II: & \left \langle \sigma v _{eff} \right  \rangle_i  n^{eq}_\chi > H, \quad \text{frezee-in} \to \text{reannihilation} \\
    II \to III: & \left \langle \sigma v _{\chi \to f} \right  \rangle_i n^{eq}_\chi > H, \quad \text{reannihilation} \to \text{freeze-out from DS} \\ \label{eq: phase transition condition 3}
    III \to IV: & \left \langle \sigma v _{\chi \to f} \right  \rangle_i  > \left \langle \sigma v _{\chi \to \phi} \right  \rangle_i , \quad \text{freeze-out from DS} \to \text{freeze-out from VS}.
\end{align}

\begin{figure}[t]
        \centering
        \includegraphics[width=0.48\linewidth]{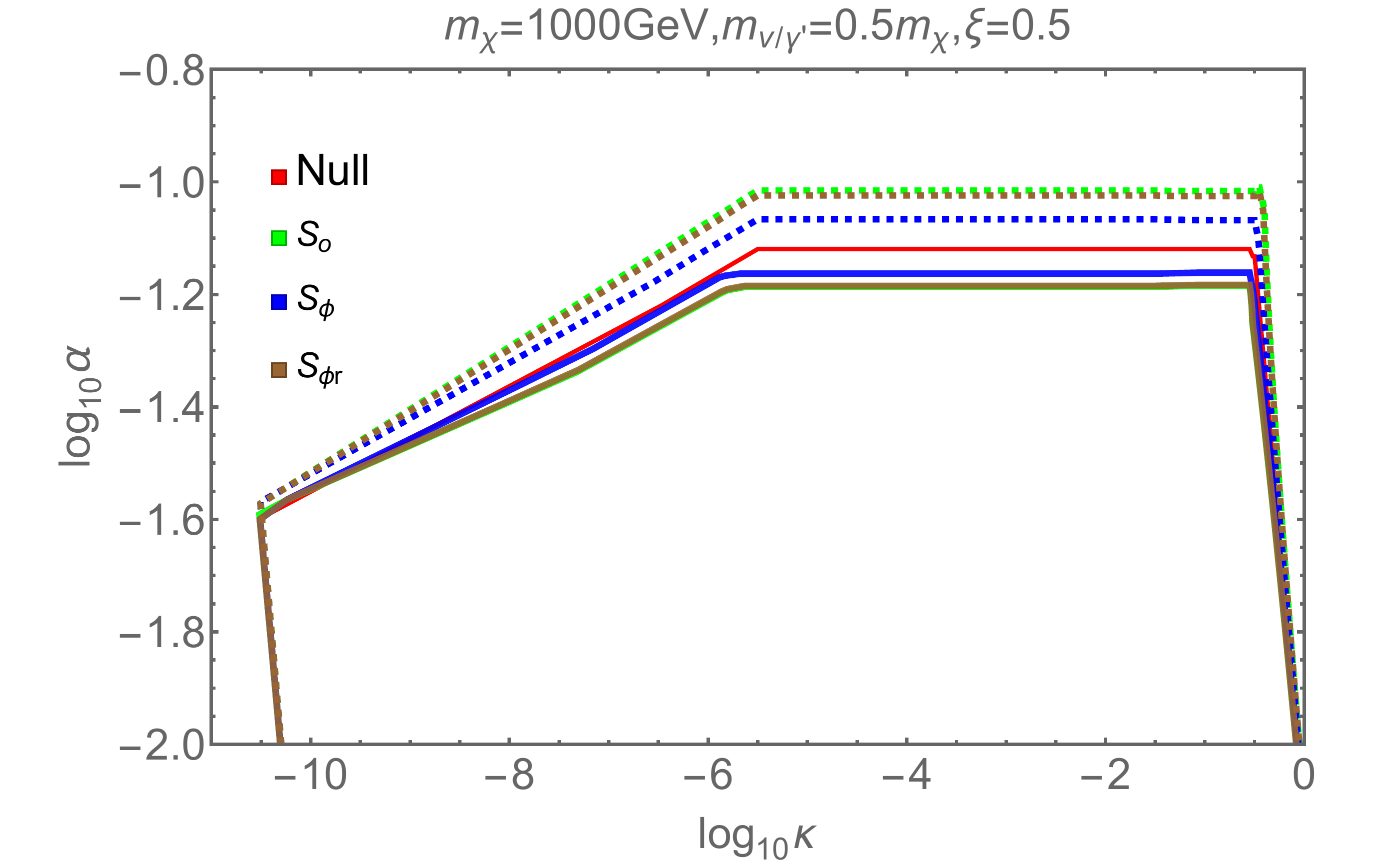}
        \includegraphics[width=0.48\linewidth]{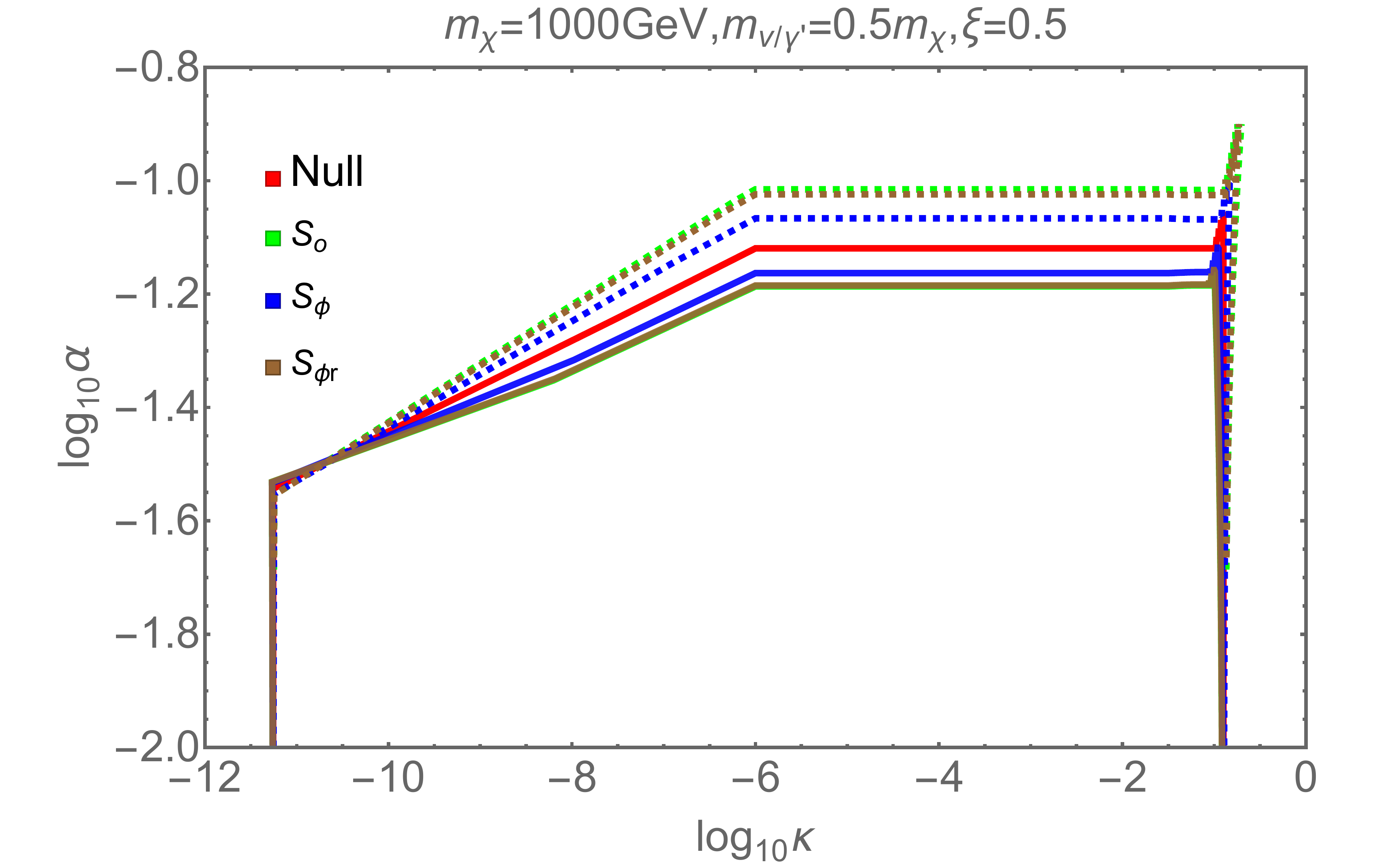}

        \includegraphics[width=0.24\linewidth]{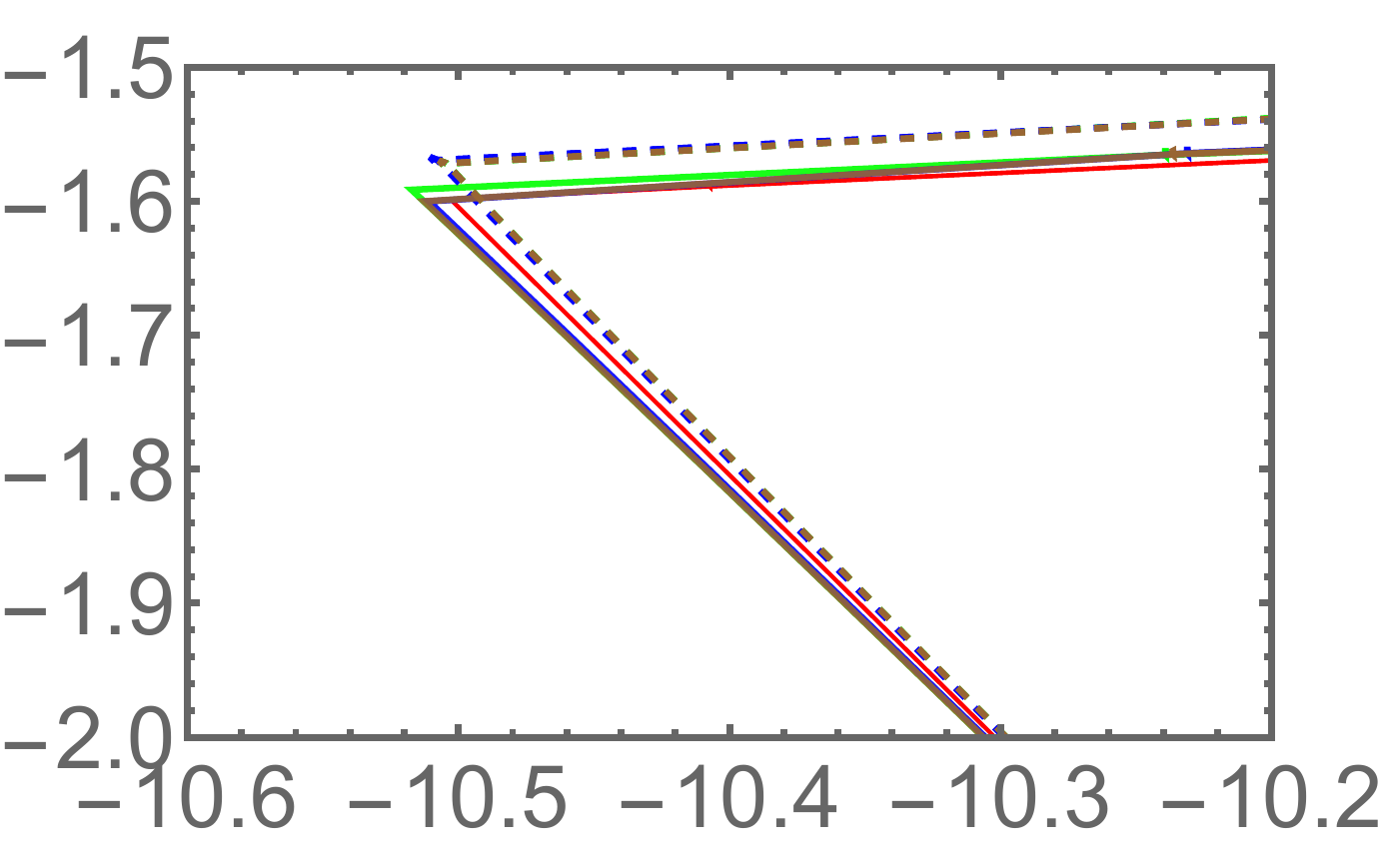}
        \includegraphics[width=0.24\linewidth]{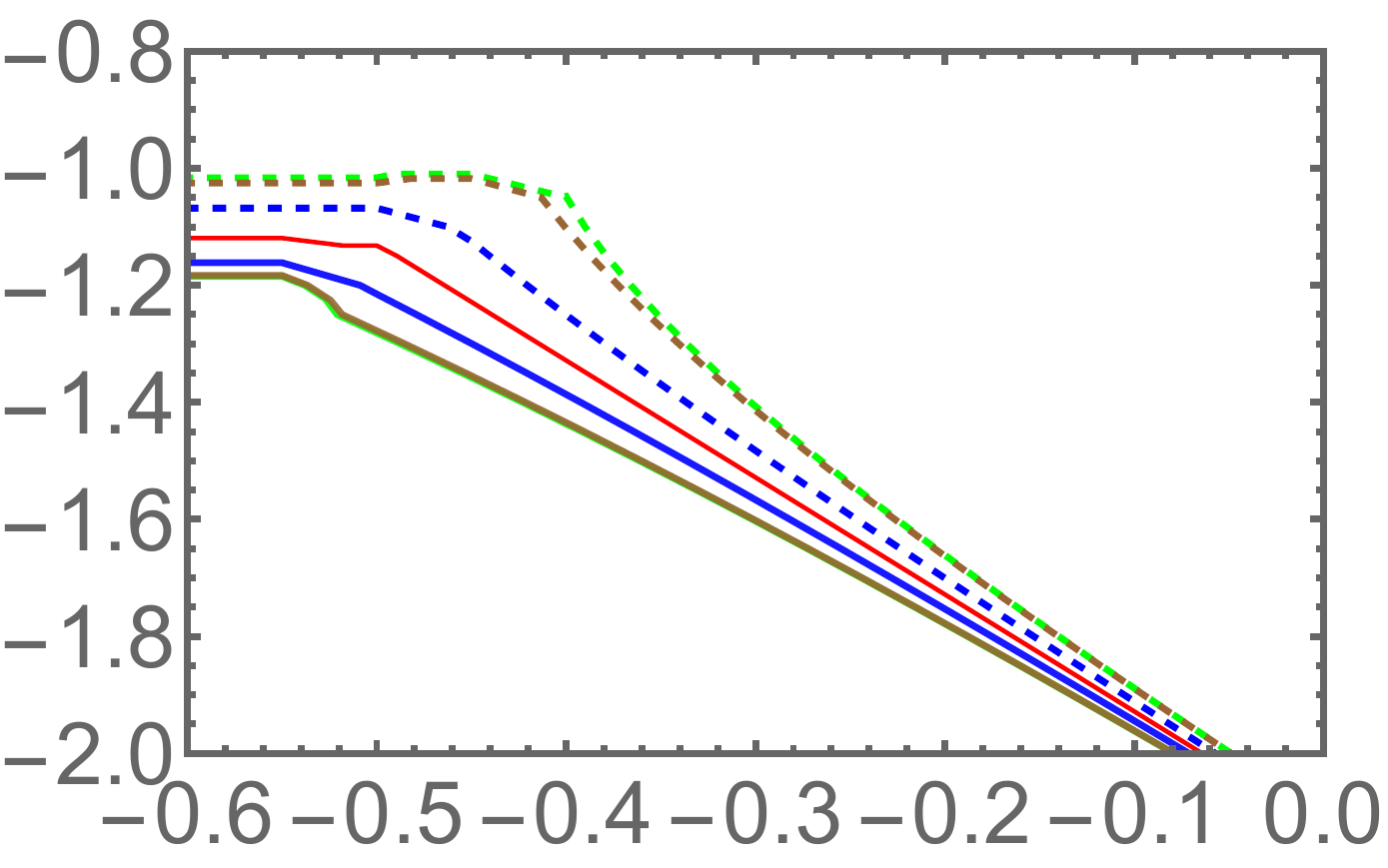}
        \includegraphics[width=0.24\linewidth]{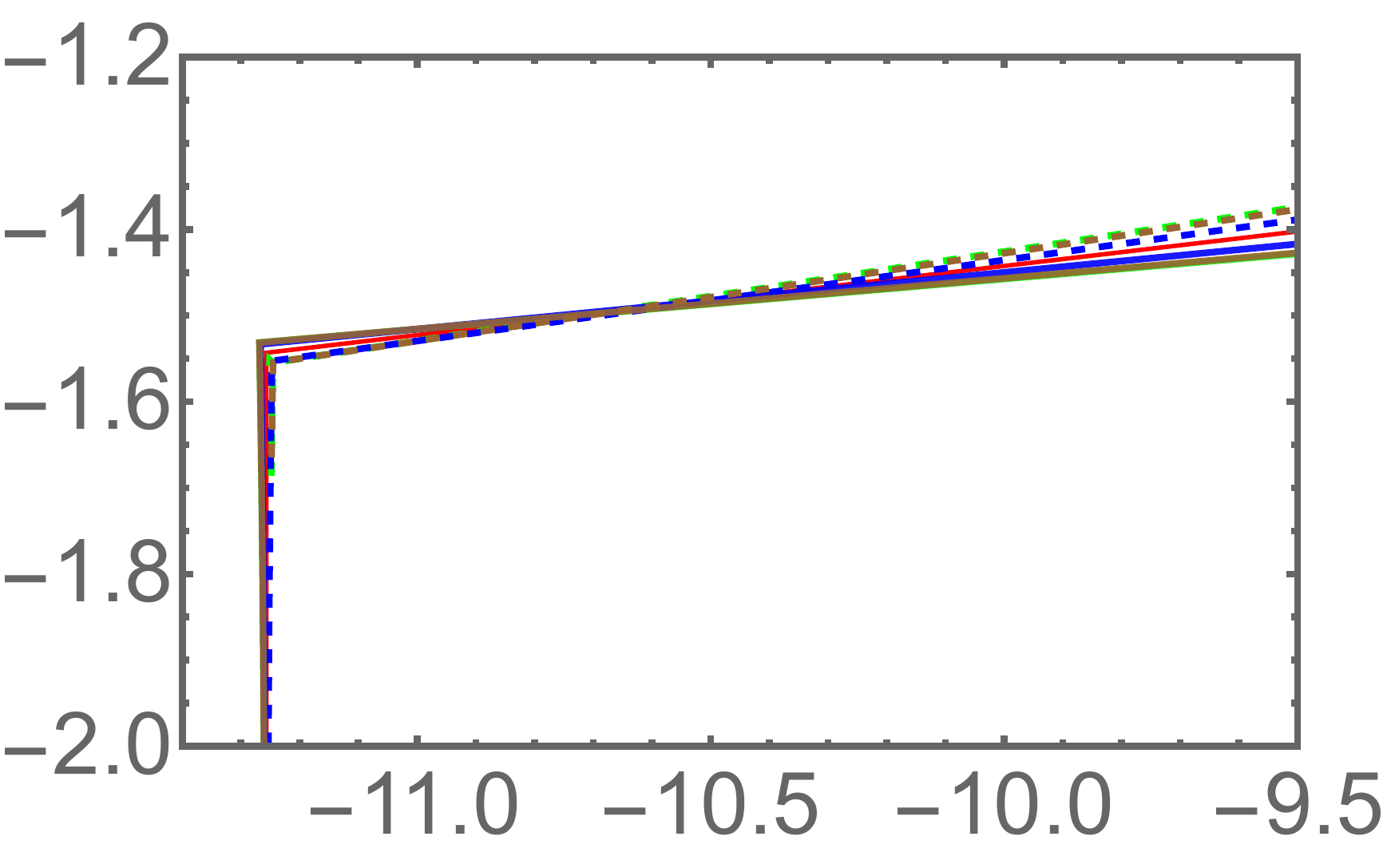}
        \includegraphics[width=0.24\linewidth]{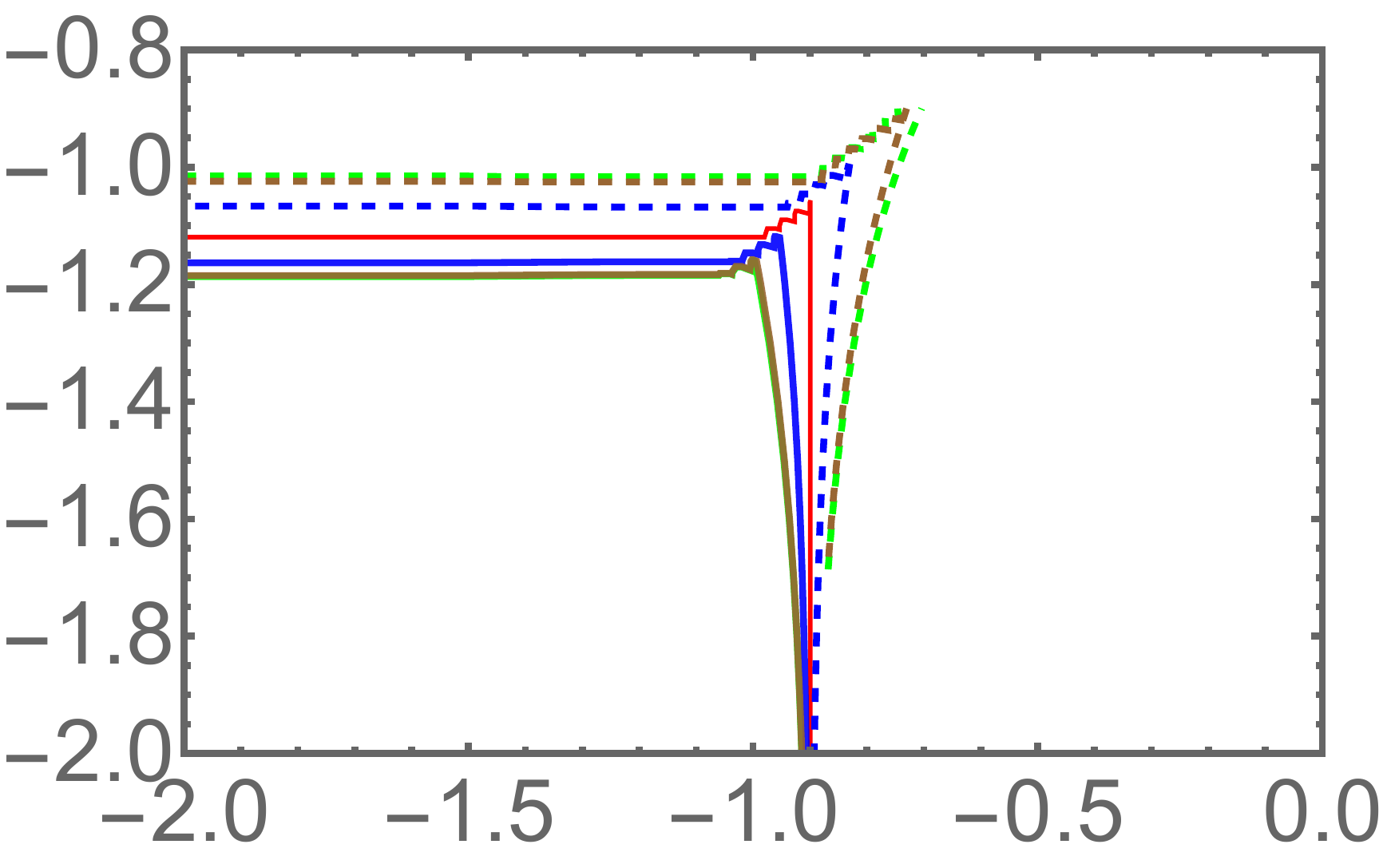}

    \caption{
    Correct relic density ($\Omega h^2 \approx 0.12$) for the U(1) dark photon models is shown in the upper left panel, while the FIMP model is presented in the upper right panel. The labels $Null, S_o, S_\phi,$ and $S_{\phi r}$ correspond to no SE, massless mediator, $0.5m_\chi$ massive mediator, and resonant mediator, respectively. Solid lines (except the red $Null$ line) represent attractive force SE, while dashed lines represent repulsive force SE. The solid green and brown lines overlap in the upper panels. To show the details of the small correction region, the zoom-in of those areas are shown in the bottom panels. Corrected parameters for the dark relic density form a band-like region. The SE has the most significant influence in the freeze-out region (right side of the upper panel), while the freeze-in region (left side of the upper panel) does not exhibit too significant changes in the possible parameter space due to the small internal coupling $\alpha$ in the DS. Note that the upper right of the FIMP panel is sharp, which may be caused by the ``hard'' phase transition condition between the DS freeze-out and the VS freeze-out (Eq \ref{eq: phase transition condition 3}).
    }\label{fig: correct omega}
\end{figure}

To obtain the present-day DM abundance, $\Omega_\chi h^2 \approx 0.12$ \cite{Planck:2018vyg}, the possible parameters space of those two models are depicted in Fig \ref{fig: correct omega} with an assumed DM mass of $1TeV$. The SE will modify the possible parameters space and result in a band-like appearance. The contours in Fig \ref{fig: correct omega} represent the portal coupling $\kappa$ and the DS internal coupling $\alpha$ parameters that produce the present-day DM abundance with the SE correction in the two specified models. The SE probably not affect the direct detection of DM, but it will affect the indirect detection, especially when considering astronomical observations that provide constraints on the DS self-interaction of $\sigma/m_\chi = 0.01 \sim 3 \ \text{cm}^2 \text{g}^{-1}$ \cite{Kahlhoefer:2015vua, Tulin:2017ara}.


The pattern of the contour lines in Fig \ref{fig: correct omega} is consistent with the discussions in sections \ref{sec: pure freeze-in} and \ref{SE in hidden}. We zoom in on the pure freeze-in region in the bottom panels of Fig \ref{fig: correct omega}. In this region, the SE does not seem significant, which is due to the small value of $\alpha$ required to achieve the correct relic density for a DM mass of $1TeV$. 
The SE effect is significant in the regions with large $\alpha$ and small $\kappa$ (top-left of the phase diagram).
Another point is that the DM relic density is primarily controlled by the coupling $\kappa$ in the pure freeze-in scenario. Even a slight change in $\kappa$ can have a significant impact on the final relic density. This is why the contour lines appear as vertical lines on the phase diagram in this region. In this region, the SE with attractive force causes a slight shift to the left of the vertical lines, while the SE with repulsive force results in a shift to the right (bottom panel of Fig \ref{fig: correct omega}) to achieve the correct relic density. The SE enhances/suppresses the cross-section, leading to a smaller/larger value of $\kappa$ compared to the standard situation without the SE, in order to match the observed relic density.

In the reannihilation region, the top-right panel shows that the attractive force SE enhancement of the cross-section will suppress the relic density (lines below the red), which means a larger value of $\kappa$ is required in the presence of the same $\alpha$ to obtain the correct relic density. Conversely, the repulsive force SE suppression leads to the opposite result. There is a crossing point in this region, which represents the enhancement and suppression switching of the relic density, as discussed in section \ref{sec: pure freeze-in}.

In the DS freeze-out region, the dominant factor controlling the relic density is the DS internal coupling $\alpha$, rather than the portal coupling $\kappa$ as in the pure freeze-in region. Increasing $\kappa$ no longer increases the relic density of DM, as the DM reaches the DS thermal equilibrium. The relic density is determined by the freeze-out process in the DS, and the freeze-out cross-section is determined by the DS internal coupling $\alpha$. The enhancement/suppression of the freeze-out cross-section will decrease/increase the DM number density during the DS freeze-out, requiring the coupling $\alpha$ smaller/larger to obtain the correct relic density (as the discussion in section \ref{sec: Freeze-out of dark sector}). That is why the contour lines (``mesa'') shift below/above under the constraint of fixed relic density. Moreover, the temperature of the DS bath is lower than the VS. The upper two factors make the SE most significant in this region. 

The SE in the usual VS freeze-out region is similar to that in the DS freeze-out region, with the main difference being that it is controlled by the coupling $\kappa$. Consistently with the previous discussion, the DM number is determined by the consumption during the freeze-out epoch. The enhancement/suppression of the cross-section will decrease/increase the DM number density, leading to a shift of the vertical line to the left/right under the constraint of a fixed relic density.

\section{Conclusion}
\label{sec:conclusion}

In this paper, we considered both the attractive and repulsive force SE under different conditions, including those with massless mediators, mediators with a mass of $0.5m_\chi$, and resonant mass mediators. Our analysis explored the performance of the SE in different phases of freeze-in. Finally, we selected two specific models, which potentially incorporate the SE, as examples to calculate the effects on the parameter space under the constraint of present-day DM relic density. We showed that using a coupling of $\alpha=0.1$ and a DM mass of $1 TeV$ is suitable and reasonable to illustrate the consequences of the SE. To avoid cumbersomeness, we avoided discussing the parameter space of the mass and coupling of the force carrier particles, as doing so would introduce an excessive number of parameters. We have covered most of the possible freeze-in phases involving the SE, but some further situations, such as the sequential freeze-in scenario, were not included in the discussion. We found that the present constraints on the DS coupling $\alpha$ and portal coupling $\kappa$ allow for a wide range of values. This implies that the parameter space is relatively safe within these constraints, resulting in a band-like structure in the possible parameter region.

We chose the coupling $\alpha$ to be approximately 0.1 to clearly present the impact of the SE on freeze-in DM in the previous discussion. To be specific, the attractive force SE will increase the DM relic density in the pure freeze-in region, resulting in a modification below $\mathcal{O}(1)$ (about 20\%) due to the relatively high temperature; while the repulsive force SE does the opposite. If the connection coupling $\kappa$ is large enough, the DM density will reach its equilibrium value, and freeze out subsequently. The attractive force SE here will make the equilibrium value larger, and subsequently freeze out to lower relic density. Similarly, the repulsive force SE does the opposite. In the ``mesa'' part of the phase diagram, the presence of attractive force SE will suppress the relic density because it enlarges the cross-section of DM with the VS/DS particles. The repulsive force SE gives the opposite results, as we have shown. The influence of the SE becomes even more significant in the DM/DS freeze-out region. In short, when the SE affects the connection reaction or the DS internal reaction, it alters the possible parameter space, resulting in a band-like possible parameter space to achieve today's DM relic density.


\section*{Acknowledgements}
Zhong thanks Dr. Dongdong Wei for the help in numerical calculation. The authors thank the anonymous reviewer for the helpful advice.

\appendix{}

\section{Kinematics}
\label{app: Kinematics}
In the center of mass (COM) frame, the incoming velocity is
\begin{align}
    v_i & = \frac{p}{\sqrt{m_1^2+p^2}} \\
    p^2 & = \frac{(s - m_1^2-m_2^2)^2 - 4m_1^2m_2^2}{4s}.
\end{align}
Take $m_1 = m_2 = m_\chi$, 
\begin{align}
    v_i = \sqrt{1-\tilde{s}^{-1}},
\end{align}
where $\tilde{s} \equiv s/4m_\chi^2$.
Take $m_3 = m_4 = m_\phi$, The outgoing velocity is
\begin{align}
    v_o = \sqrt{1- z \ \tilde{s}^{-1}},
\end{align}
where $z = m_\phi^2/m_\chi^2$. The cross-section for the $2 \to 2$ situation is 
\begin{align}
    \frac{d \sigma}{d \Omega} & =  \frac{1}{64 \pi^2 s} \frac{\left|\vec{p}_f\right|}{\left|\vec{p}_i\right|} |\mathcal{M}|^2 \Theta ( s-4m_\phi^2 ) \\
    & =   \frac{1}{256 m_\chi^2 \pi^2 \tilde{s}} \frac{\sqrt{1- z \ \tilde{s}^{-1}}}{\sqrt{1-  \ \tilde{s}^{-1}}} \Theta (\tilde{s} - z) \Theta (\tilde{s} - 1) |\mathcal{M}|^2.
\end{align}
The total cross-section is 
\begin{align}
    \sigma(\tilde{s}) = \frac{1}{256\pi^2 \ m_\chi^2  \ \tilde{s}} \frac{\sqrt{1- z \tilde{s}^{-1}}}{\sqrt{1- \tilde{s}^{-1}}} \Theta (\tilde{s} - z) \Theta (\tilde{s} - 1) \int d \Omega |\mathcal{M}|^2.
\end{align}
The Mandelstam varables are
\begin{align}
    t/m_\chi^2 & =  1 + z - 2\tilde{s} \ [1 - \sqrt{(1- \tilde{s}^{-1})(1- z \tilde{s}^{-1})} \cos \theta]\\
    u/m_\chi^2 & =  1 + z + 2\tilde{s} \ [1 + \sqrt{(1- \tilde{s}^{-1})(1- z \tilde{s}^{-1})} \cos \theta].
\end{align}

\section{The upper bound of pure freeze-in}
\label{app: upper bound}
Assuming $g_{*s}$ is constant
\begin{align}
    Y_\chi & =  \int_{0}^{\infty}  dx \frac{xs}{H(m_\chi)}(1+\frac{T}{3g_{*s}}\frac{dg_{*s}}{dT}) \cdot \left \langle \sigma v _{\chi \chi \to f f } \right  \rangle_{eh} \cdot Y^{eq \ 2}_{\chi} \\
    & = C_1 \int_{\tilde{s}_{min}}^{\infty} d \tilde{s} \sqrt{\tilde{s}} \ (\tilde{s}-1) \   \sigma_{eh}(\tilde{s}) \int_{0}^{\infty} dx \ (x^3) K_1(2 \sqrt{\tilde{s}} x) \\
    & = C_1 \int_{\tilde{s}_{min}}^{\infty} d \tilde{s} \ \frac{\tilde{s}-1}{\tilde{s}^{3/2}} \cdot \sigma_{i}(\tilde{s}),
\end{align}
where 
\begin{align}
    C_1 & = \frac{405 \sqrt{5} g^2 m_{pl} m_{\chi }}{128 \pi ^{13/2} g_{*s} g_*^{1/2}}.
\end{align}

\section{The $\alpha$ threshold on pure freeze-in}
\label{app: alpha threshold}
To satisfy the pure freeze-in condition, it is required that the DM particle $\chi$ and its mediator $\phi$ are not in thermal equilibrium with the SM bath. Assuming the cross-sections $SM \to \chi$ and $\chi \to \phi$ are 
\begin{align}
    & \left \langle \sigma_{f \to \chi} v \right \rangle \sim \frac{\kappa^2}{4 m_\chi^2 \tilde s} \\
    & \left \langle \sigma_{\chi \to \phi} \right \rangle \sim \frac{\alpha^2}{4 m_\chi^2 \tilde s}.
\end{align}
The equilibrium density of $\chi$ and the Hubble rate are
\begin{align}
    & n^{eq}_{\chi} = \frac{g \ \pi^2}{30} (\frac{m_\chi}{x})^3 \\
    & H = \sqrt{\frac{4 \pi^3 g_* }{45}} \frac{m_\chi^2}{x_f^2 \ m_{pl}}.
\end{align}
Solving the equation  
\begin{align}
    \sqrt{\left \langle \sigma_{f \to \chi} v \right \rangle \left \langle \sigma_{\chi \to \phi} \right \rangle } \ n^{eq}_{\chi}  = H,
\end{align}
we obtain the threshold of pure freeze-in as 
\begin{align} \label{Eq. ath}
    \alpha_{th} = \sqrt{\frac{g_*}{5 \pi}} \frac{80 \ x_f \ \tilde s \ m_\chi}{\ g \ \kappa \ m_{pl}}.
\end{align}
Taking $x_f = 2, \ \tilde s = 2, \ \kappa =10^{-12}, \ g=1, \ m_\chi = 1TeV$, we find that $\alpha \approx 0.07$, which is approximately 0.1. This is a conservative estimate as we assume that the $\chi$ particles are in equilibrium with the SM bath. 

Another estimation is provided in Eq.(3.21) of \cite{Hambye:2019dwd}. We start by calculating the threshold for the $\chi$ particle in the SM bath:
\begin{align}
     \left \langle \sigma_{f \to \chi} v \right \rangle n^{eq}_{\chi} = H.
\end{align}
This gives us the threshold for $\kappa$ as
\begin{align}
    \kappa^2_{th} = \sqrt{\frac{g_*}{5 \pi}} \frac{80 \ x_f \ \tilde s \ m_\chi}{g \ m_{pl}}.
\end{align}
The number density of $\chi$ can be estimated as
\begin{align}
    n_\chi = \frac{\kappa^2}{\kappa^2_{th}} \ n^{eq}_\chi.
\end{align}
Now make sure that the reaction rate of $\phi$ is smaller than $H$
\begin{align}
    \left \langle \sigma_{\chi \to \phi} \right \rangle n_\chi = H,
\end{align}
the threshold of $\alpha$ is
\begin{align}
    \alpha_{th} = \sqrt{\frac{g_*}{5 \pi}} \frac{80 \ x_f \ \tilde s \ m_\chi}{\ g \ \kappa \ m_{pl}},
\end{align}
which is the same as Eq. \ref{Eq. ath}.

\section{The Boltzmann equation with DS}
\label{app: Bol with DS}
The DS contains particles $\chi$ and $\phi$, and the Boltzmann equations are 
\begin{align}
    & \frac{d n_\chi}{dt} + 3 H n_\chi = \left \langle \sigma v _{\chi \to  f } \right  \rangle \cdot (n_\chi ^{eq \ 2} - n_\chi^2) - \left \langle \sigma v _{\chi \to  \phi } \right  \rangle \cdot  n_\chi ^2 + \left \langle \sigma v _{ \phi \to \chi } \right  \rangle \cdot n_{\phi}^2 \\
    & \frac{d n_\phi}{dt} + 3 H n_{\phi} = \left \langle \sigma v _{\chi \to  \phi } \right  \rangle \cdot  n_\chi ^2 - \left \langle \sigma v _{ \phi \to \chi } \right  \rangle \cdot n_{\phi}^2,
\end{align}
where $f$ is in the equilibrium of VS bath. Using the total entropy conserve condition $d(a^3 s) = 0$, equations are reduced as below
\begin{align}\label{eq: Bolzman 2}
        & \frac{\dot Y_\chi}{s}  = \left \langle \sigma v _{\chi \to  f } \right  \rangle \cdot (Y_\chi ^{eq \ 2} - Y_\chi^2) - \left \langle \sigma v _{\chi \to  \phi} \right  \rangle \cdot  Y_\chi ^2 + \left \langle \sigma v _{ \phi \to \chi } \right  \rangle \cdot Y_{\phi}^2 \\
    & \frac{\dot Y_{\phi}}{s} = \left \langle \sigma v _{\chi \to  \phi} \right  \rangle \cdot  Y_\chi ^2 - \left \langle \sigma v _{ \phi \to \chi } \right  \rangle \cdot Y_{\phi}^2.
\end{align}
Assuming VS and DS with the energy density $\rho, \rho'$ and temperature $T, T'$. In the early time, we do not consider vacuum energy. The Friedmann equation is
\begin{align}
      H^2 & = \frac{8 \pi G}{3} \ (\rho+\rho')  = \frac{4 \pi^3 G}{45} \ g_{*eff} \ T^4.
 \end{align}
We define the effective relativistic degrees of freedom as
\begin{align}
    g_{*eff} = g_* + g_{*'}  \xi^4.
\end{align}
The total entropy density is  
\begin{align}
    s & = \frac{2 \pi^3}{45} (g_{*s}+g_{*s'} \xi^3)T^3 = \frac{2 \pi^3}{45} g_{seff} T^3,
\end{align}
where the effective entropy degrees of freedom is defined as 
\begin{align}
    g_{seff} = g_{*s} + g_{*s'} \xi^3.
\end{align}
In the radiation-dominated epoch, $\rho \propto a^{-4}$. Then we get the relationship between time and temperature
\begin{align}
    t = (\frac{45}{16 \pi^3})^{1/2}  \frac{m_{pl}}{g_{*eff}^{1/2} \ T^2}.
\end{align}
The derivative of $t$ with respect to $T$ is
\begin{align}
    \frac{dt}{dT} = -\frac{x^2}{H(m_\chi)}(\frac{1}{T} + \frac{1}{4g_{*eff}} \frac{dg_{*eff}}{dT}).
\end{align}
Substitute $t$ with $x$ 
\begin{align}
    \frac{d}{dt} =  ( \frac{dt}{dT} ) ^{-1} \frac{dx}{dT} \frac{d}{dx}.
\end{align}
Therefore, the Boltzmann equations Eq. \ref{eq: Bolzman 2} can be rewritten as
\begin{align}
    & \frac{dY_\chi}{dx} = \frac{xs}{H_{m_{\chi}}}(1+\frac{T}{4 g_{*eff}} \frac{d g_{*eff}}{dT})[ \left \langle \sigma v _{\chi \to  f } \right  \rangle \cdot (Y_\chi ^{eq \ 2} - Y_\chi^2) - \left \langle \sigma v _{\chi \to  \phi} \right  \rangle \cdot  Y_\chi ^2 + \left \langle \sigma v _{ \phi \to \chi } \right  \rangle \cdot Y_{\phi}^2] \\
    & \frac{dY_{\phi}}{dx} = \frac{xs}{H_{m_{\chi}}}(1+\frac{T}{4 g_{*eff}} \frac{d g_{*eff}}{dT}) [\left \langle \sigma v _{\chi \to  \phi } \right  \rangle \cdot  Y_\chi ^2 - \left \langle \sigma v _{\phi \to \chi } \right  \rangle \cdot Y_{\phi}^2].
\end{align}
We encode the VS and DS temperature relationship into the effect on relativistic degrees of freedom $g_{*eff}$ and $g_{seff}$. If DS entropy and relativistic degrees of freedom are small, indicating $\xi < 1$, it is safe to ignore the contribution of the DS.

Assuming that $\phi$ is in thermal equilibrium with the DS bath at a temperature $T'$, the second Boltzmann equation about $n_{\phi}$ can be eliminated, and the first Boltzmann equation is reduced as 
\begin{align}
    & \frac{H_{m_{\chi}}}{xs} \frac{dY_\chi}{dx} =  \left \langle \sigma v _{\chi \to  f } \right  \rangle \cdot (Y_\chi ^{eq \ 2} - Y_\chi^2) + {\left \langle \sigma v _{\chi \to  \phi } \right  \rangle }  \cdot  ( {Y'}_\chi^{eq \ 2} - {Y}_\chi ^2).
\end{align}
If the coupling between dark sector particles is much larger than the coupling between the dark and visual sector particles, it gives
\begin{align}
    & \left \langle \sigma v _{\chi \to  f } \right  \rangle << \left \langle \sigma v _{\chi \to  \phi } \right  \rangle.
\end{align}
Finally, the Boltzmann equation can be expressed as
\begin{align}
    \frac{H_{m_{\chi}}}{xs} \frac{dY_\chi}{dx} =  \left \langle \sigma v _{\chi \to  f } \right  \rangle \cdot Y_\chi ^{eq \ 2}  + \left \langle \sigma v _{\chi \to  \phi } \right  \rangle \cdot  ( {Y'}_\chi^{eq \ 2} - Y_\chi ^2),
\end{align}
where  
\begin{align}
     {Y'}_\chi^{eq}  = \xi^3 \ Y_\chi^{eq}.
\end{align}

\section{Balance condition}
\label{app: balance condition}
Note that for the freeze-in process, we have
\begin{align} 
    &  \left \langle \sigma v _{f \to \chi } \right  \rangle n_f^{eq} > H(T) \\
    & \left \langle \sigma v _{\chi \to f } \right  \rangle n_\chi < H(T).
\end{align}
Furthermore, considering the detailed balance condition:
\begin{align}
    \left \langle \sigma v _{f \to \chi } \right  \rangle n_f^{eq} = \left \langle \sigma v _{\chi \to f } \right  \rangle n_\chi^{eq},
\end{align}
we have 
\begin{align}
     \frac{n_\chi}{n_\chi^{eq}} = \delta < 1,
\end{align}
replaced by the 
\begin{align}
    & \delta  = \xi^3 , \ \text{radiation }\chi\\
    & \delta = \frac{ (x \xi)^{3/2} }{ e^{-x/\xi }}, \ \text{matter }\chi .
\end{align}
For the radiation $\chi$, we have
\begin{align}
    & 0 < \xi < 1.
\end{align}
For the matter $\chi$, its minimum value is
\begin{align}
    \delta_{min} =  (\frac{2e}{3})^{3/2} x^3.
\end{align}
Considering $x \sim 1 $, it is impossible to have a proper non-relativistic thermodynamic description with the temperature $T'$. In other words, the contribution to the consumption of $\chi$ is negligible, which is consistent with Boltzmann suppression. Therefore, we take 
\begin{align}
    \xi \simeq  0 , \  \delta \simeq  0.
\end{align}

Next, considering 
\begin{align}
      \left \langle \sigma v _{\chi \to f } \right  \rangle n_\chi^{eq} = H(T),
\end{align}
we take $n_\chi^{eq} \sim (m_\chi/x)^3, g_* =100$ and $x \sim 1$ at the freeze-in and obtain
\begin{align}
    \kappa^2 \sim \sqrt{\frac{4 \pi^3 g_*}{45}} \frac{x m_\chi} {m_{pl}} \sim \frac{10 m_\chi}{m_{pl}}.
\end{align}

\bibliography{ref}

\providecommand{\href}[2]{#2}\begingroup\raggedright\begin{thebibliography}{10}

\bibitem{https://doi.org/10.1002/andp.19314030302}
A.~Sommerfeld, \emph{Über die beugung und bremsung der elektronen},
  \href{https://doi.org/https://doi.org/10.1002/andp.19314030302}{\emph{Annalen
  der Physik} {\bfseries 403} (1931) 257}
  [\href{https://arxiv.org/abs/https://onlinelibrary.wiley.com/doi/pdf/10.1002/andp.19314030302}{{\ttfamily
  https://onlinelibrary.wiley.com/doi/pdf/10.1002/andp.19314030302}}].

\bibitem{Lee:1977ua}
B.W.~Lee and S.~Weinberg, \emph{{Cosmological Lower Bound on Heavy Neutrino
  Masses}}, \href{https://doi.org/10.1103/PhysRevLett.39.165}{\emph{Phys. Rev.
  Lett.} {\bfseries 39} (1977) 165}.

\bibitem{Kolb:1990vq}
E.W.~Kolb and M.S.~Turner, \emph{{The Early Universe}}, vol.~69 (1990),
  \href{https://doi.org/10.1201/9780429492860}{10.1201/9780429492860}.

\bibitem{Du:2021jcj}
Y.~Du, F.~Huang, H.-L.~Li, Y.-Z.~Li and J.-H.~Yu, \emph{{Revisiting dark matter
  freeze-in and freeze-out through phase-space distribution}},
  \href{https://doi.org/10.1088/1475-7516/2022/04/012}{\emph{JCAP} {\bfseries
  04} (2022) 012} [\href{https://arxiv.org/abs/2111.01267}{{\ttfamily
  2111.01267}}].

\bibitem{Hall:2009bx}
L.J.~Hall, K.~Jedamzik, J.~March-Russell and S.M.~West, \emph{{Freeze-In
  Production of FIMP Dark Matter}},
  \href{https://doi.org/10.1007/JHEP03(2010)080}{\emph{JHEP} {\bfseries 03}
  (2010) 080} [\href{https://arxiv.org/abs/0911.1120}{{\ttfamily 0911.1120}}].

\bibitem{Bernal:2017kxu}
N.~Bernal, M.~Heikinheimo, T.~Tenkanen, K.~Tuominen and V.~Vaskonen, \emph{{The
  Dawn of FIMP Dark Matter: A Review of Models and Constraints}},
  \href{https://doi.org/10.1142/S0217751X1730023X}{\emph{Int. J. Mod. Phys. A}
  {\bfseries 32} (2017) 1730023}
  [\href{https://arxiv.org/abs/1706.07442}{{\ttfamily 1706.07442}}].

\bibitem{Elahi:2014fsa}
F.~Elahi, C.~Kolda and J.~Unwin, \emph{{UltraViolet Freeze-in}},
  \href{https://doi.org/10.1007/JHEP03(2015)048}{\emph{JHEP} {\bfseries 03}
  (2015) 048} [\href{https://arxiv.org/abs/1410.6157}{{\ttfamily 1410.6157}}].

\bibitem{McDonald:2001vt}
J.~McDonald, \emph{{Thermally generated gauge singlet scalars as
  selfinteracting dark matter}},
  \href{https://doi.org/10.1103/PhysRevLett.88.091304}{\emph{Phys. Rev. Lett.}
  {\bfseries 88} (2002) 091304}
  [\href{https://arxiv.org/abs/hep-ph/0106249}{{\ttfamily hep-ph/0106249}}].

\bibitem{vonHarling:2014kha}
B.~von Harling and K.~Petraki, \emph{{Bound-state formation for thermal relic
  dark matter and unitarity}},
  \href{https://doi.org/10.1088/1475-7516/2014/12/033}{\emph{JCAP} {\bfseries
  12} (2014) 033} [\href{https://arxiv.org/abs/1407.7874}{{\ttfamily
  1407.7874}}].

\bibitem{2015PhR...555....1B}
H.~{Baer}, K.-Y.~{Choi}, J.E.~{Kim} and L.~{Roszkowski}, \emph{{Dark matter
  production in the early Universe: Beyond the thermal WIMP paradigm}},
  \href{https://doi.org/10.1016/j.physrep.2014.10.002}{\emph{physrep}
  {\bfseries 555} (2015) 1} [\href{https://arxiv.org/abs/1407.0017}{{\ttfamily
  1407.0017}}].

\bibitem{Feng:2010zp}
J.L.~Feng, M.~Kaplinghat and H.-B.~Yu, \emph{{Sommerfeld Enhancements for
  Thermal Relic Dark Matter}},
  \href{https://doi.org/10.1103/PhysRevD.82.083525}{\emph{Phys. Rev. D}
  {\bfseries 82} (2010) 083525}
  [\href{https://arxiv.org/abs/1005.4678}{{\ttfamily 1005.4678}}].

\bibitem{Arkani-Hamed:2008hhe}
N.~Arkani-Hamed, D.P.~Finkbeiner, T.R.~Slatyer and N.~Weiner, \emph{{A Theory
  of Dark Matter}},
  \href{https://doi.org/10.1103/PhysRevD.79.015014}{\emph{Phys. Rev. D}
  {\bfseries 79} (2009) 015014}
  [\href{https://arxiv.org/abs/0810.0713}{{\ttfamily 0810.0713}}].

\bibitem{Hisano:2004ds}
J.~Hisano, S.~Matsumoto, M.M.~Nojiri and O.~Saito, \emph{{Non-perturbative
  effect on dark matter annihilation and gamma ray signature from galactic
  center}}, \href{https://doi.org/10.1103/PhysRevD.71.063528}{\emph{Phys. Rev.
  D} {\bfseries 71} (2005) 063528}
  [\href{https://arxiv.org/abs/hep-ph/0412403}{{\ttfamily hep-ph/0412403}}].

\bibitem{Ellis:2018jyl}
J.~Ellis, J.L.~Evans, F.~Luo, K.A.~Olive and J.~Zheng, \emph{{Stop
  Coannihilation in the CMSSM and SubGUT Models}},
  \href{https://doi.org/10.1140/epjc/s10052-018-5831-z}{\emph{Eur. Phys. J. C}
  {\bfseries 78} (2018) 425}
  [\href{https://arxiv.org/abs/1801.09855}{{\ttfamily 1801.09855}}].

\bibitem{Ellis:2015vna}
J.~Ellis, J.L.~Evans, F.~Luo and K.A.~Olive, \emph{{Scenarios for Gluino
  Coannihilation}}, \href{https://doi.org/10.1007/JHEP02(2016)071}{\emph{JHEP}
  {\bfseries 02} (2016) 071}
  [\href{https://arxiv.org/abs/1510.03498}{{\ttfamily 1510.03498}}].

\bibitem{Wang:2022avs}
X.~Wang, F.~Zhong and F.~Luo, \emph{{Final bound-state formation effect on dark
  matter annihilation}},  \href{https://arxiv.org/abs/2204.01091}{{\ttfamily
  2204.01091}}.

\bibitem{Enqvist:1992va}
K.~Enqvist, K.~Kainulainen and I.~Vilja, \emph{{Phase transitions in the
  singlet majoron model}},
  \href{https://doi.org/10.1016/0550-3213(93)90369-Z}{\emph{Nucl. Phys. B}
  {\bfseries 403} (1993) 749}.

\bibitem{Enqvist:2014zqa}
K.~Enqvist, S.~Nurmi, T.~Tenkanen and K.~Tuominen, \emph{{Standard Model with a
  real singlet scalar and inflation}},
  \href{https://doi.org/10.1088/1475-7516/2014/08/035}{\emph{JCAP} {\bfseries
  08} (2014) 035} [\href{https://arxiv.org/abs/1407.0659}{{\ttfamily
  1407.0659}}].

\bibitem{Planck:2018vyg}
{\scshape Planck} collaboration, \emph{{Planck 2018 results. VI. Cosmological
  parameters}},
  \href{https://doi.org/10.1051/0004-6361/201833910}{\emph{Astron. Astrophys.}
  {\bfseries 641} (2020) A6}
  [\href{https://arxiv.org/abs/1807.06209}{{\ttfamily 1807.06209}}].

\bibitem{2009JHEP...05..024I}
R.~{Iengo}, \emph{{Sommerfeld enhancement: general results from field theory
  diagrams}},
  \href{https://doi.org/10.1088/1126-6708/2009/05/024}{\emph{Journal of High
  Energy Physics} {\bfseries 2009} (2009) 024}
  [\href{https://arxiv.org/abs/0902.0688}{{\ttfamily 0902.0688}}].

\bibitem{Slatyer:2009vg}
T.R.~Slatyer, \emph{{The Sommerfeld enhancement for dark matter with an excited
  state}}, \href{https://doi.org/10.1088/1475-7516/2010/02/028}{\emph{JCAP}
  {\bfseries 02} (2010) 028} [\href{https://arxiv.org/abs/0910.5713}{{\ttfamily
  0910.5713}}].

\bibitem{2010JPhG...37j5009C}
S.~{Cassel}, \emph{{Sommerfeld factor for arbitrary partial wave processes}},
  \href{https://doi.org/10.1088/0954-3899/37/10/105009}{\emph{Journal of
  Physics G Nuclear Physics} {\bfseries 37} (2010) 105009}
  [\href{https://arxiv.org/abs/0903.5307}{{\ttfamily 0903.5307}}].

\bibitem{Elor:2018xku}
G.~Elor, H.~Liu, T.R.~Slatyer and Y.~Soreq, \emph{{Complementarity for Dark
  Sector Bound States}},
  \href{https://doi.org/10.1103/PhysRevD.98.036015}{\emph{Phys. Rev. D}
  {\bfseries 98} (2018) 036015}
  [\href{https://arxiv.org/abs/1801.07723}{{\ttfamily 1801.07723}}].

\bibitem{Hisano:2003ec}
J.~Hisano, S.~Matsumoto and M.M.~Nojiri, \emph{{Explosive dark matter
  annihilation}},
  \href{https://doi.org/10.1103/PhysRevLett.92.031303}{\emph{Phys. Rev. Lett.}
  {\bfseries 92} (2004) 031303}
  [\href{https://arxiv.org/abs/hep-ph/0307216}{{\ttfamily hep-ph/0307216}}].

\bibitem{2017JCAP...08..003K}
F.~{Kahlhoefer}, K.~{Schmidt-Hoberg} and S.~{Wild}, \emph{{Dark matter
  self-interactions from a general spin-0 mediator}},
  \href{https://doi.org/10.1088/1475-7516/2017/08/003}{\emph{jcap} {\bfseries
  2017} (2017) 003} [\href{https://arxiv.org/abs/1704.02149}{{\ttfamily
  1704.02149}}].

\bibitem{2019PhRvD..99c6016T}
Y.-L.~{Tang} and G.-L.~{Zhou}, \emph{{Calculations of the Sommerfeld effect in
  a unified wave function framework}},
  \href{https://doi.org/10.1103/PhysRevD.99.036016}{\emph{prd} {\bfseries 99}
  (2019) 036016} [\href{https://arxiv.org/abs/1806.10124}{{\ttfamily
  1806.10124}}].

\bibitem{1991NuPhB.360..145G}
P.~{Gondolo} and G.~{Gelmini}, \emph{{Cosmic abundances of stable particles:
  improved analysis.}},
  \href{https://doi.org/10.1016/0550-3213(91)90438-4}{\emph{Nuclear Physics B}
  {\bfseries 360} (1991) 145}.

\bibitem{Bringmann:2021sth}
T.~Bringmann, S.~Heeba, F.~Kahlhoefer and K.~Vangsnes, \emph{{Freezing-in a hot
  bath: resonances, medium effects and phase transitions}},
  \href{https://doi.org/10.1007/JHEP02(2022)110}{\emph{JHEP} {\bfseries 02}
  (2022) 110} [\href{https://arxiv.org/abs/2111.14871}{{\ttfamily
  2111.14871}}].

\bibitem{DAgnolo:2015ujb}
R.T.~D'Agnolo and J.T.~Ruderman, \emph{{Light Dark Matter from Forbidden
  Channels}}, \href{https://doi.org/10.1103/PhysRevLett.115.061301}{\emph{Phys.
  Rev. Lett.} {\bfseries 115} (2015) 061301}
  [\href{https://arxiv.org/abs/1505.07107}{{\ttfamily 1505.07107}}].

\bibitem{Cui:2020ppc}
X.~Cui and F.~Luo, \emph{{Final state Sommerfeld effect on dark matter relic
  abundance}}, \href{https://doi.org/10.1007/JHEP01(2021)156}{\emph{JHEP}
  {\bfseries 01} (2021) 156}
  [\href{https://arxiv.org/abs/2009.14591}{{\ttfamily 2009.14591}}].

\bibitem{Biswas:2022vkq}
A.~Biswas, D.~Borah, N.~Das and D.~Nanda, \emph{{Freeze-in Dark Matter and
  $\Delta N_{eff}$ via Light Dirac Neutrino Portal}},
  \href{https://arxiv.org/abs/2205.01144}{{\ttfamily 2205.01144}}.

\bibitem{Chu:2011be}
X.~Chu, T.~Hambye and M.H.G.~Tytgat, \emph{{The Four Basic Ways of Creating
  Dark Matter Through a Portal}},
  \href{https://doi.org/10.1088/1475-7516/2012/05/034}{\emph{JCAP} {\bfseries
  05} (2012) 034} [\href{https://arxiv.org/abs/1112.0493}{{\ttfamily
  1112.0493}}].

\bibitem{Bharucha:2022lty}
A.~Bharucha, F.~Br\"ummer, N.~Desai and S.~Mutzel, \emph{{Axion-like particles
  as mediators for dark matter: beyond freeze-out}},
  \href{https://arxiv.org/abs/2209.03932}{{\ttfamily 2209.03932}}.

\bibitem{Berbig:2022nre}
M.~Berbig, \emph{{Freeze-In of radiative keV-scale neutrino dark matter from a
  new U(1)$_{B-L}$}},
  \href{https://doi.org/10.1007/JHEP09(2022)101}{\emph{JHEP} {\bfseries 09}
  (2022) 101} [\href{https://arxiv.org/abs/2203.04276}{{\ttfamily
  2203.04276}}].

\bibitem{Luo:2020fdt}
X.~Luo, W.~Rodejohann and X.-J.~Xu, \emph{{Dirac neutrinos and N$_{eff}$. Part
  II. The freeze-in case}},
  \href{https://doi.org/10.1088/1475-7516/2021/03/082}{\emph{JCAP} {\bfseries
  03} (2021) 082} [\href{https://arxiv.org/abs/2011.13059}{{\ttfamily
  2011.13059}}].

\bibitem{DEramo:2020gpr}
F.~D'Eramo and A.~Lenoci, \emph{{Lower mass bounds on FIMP dark matter produced
  via freeze-in}},
  \href{https://doi.org/10.1088/1475-7516/2021/10/045}{\emph{JCAP} {\bfseries
  10} (2021) 045} [\href{https://arxiv.org/abs/2012.01446}{{\ttfamily
  2012.01446}}].

\bibitem{Matsui:2015maa}
H.~Matsui and M.~Nojiri, \emph{{Higgs sector extension of the neutrino minimal
  standard model with thermal freeze-in production mechanism}},
  \href{https://doi.org/10.1103/PhysRevD.92.025045}{\emph{Phys. Rev. D}
  {\bfseries 92} (2015) 025045}
  [\href{https://arxiv.org/abs/1503.01293}{{\ttfamily 1503.01293}}].

\bibitem{Bae:2017dpt}
K.J.~Bae, A.~Kamada, S.P.~Liew and K.~Yanagi, \emph{{Light axinos from
  freeze-in: production processes, phase space distributions, and Ly-$\alpha$
  forest constraints}},
  \href{https://doi.org/10.1088/1475-7516/2018/01/054}{\emph{JCAP} {\bfseries
  01} (2018) 054} [\href{https://arxiv.org/abs/1707.06418}{{\ttfamily
  1707.06418}}].

\bibitem{DEramo:2017ecx}
F.~D'Eramo, N.~Fernandez and S.~Profumo, \emph{{Dark Matter Freeze-in
  Production in Fast-Expanding Universes}},
  \href{https://doi.org/10.1088/1475-7516/2018/02/046}{\emph{JCAP} {\bfseries
  02} (2018) 046} [\href{https://arxiv.org/abs/1712.07453}{{\ttfamily
  1712.07453}}].

\bibitem{Hambye:2019dwd}
T.~Hambye, M.H.G.~Tytgat, J.~Vandecasteele and L.~Vanderheyden, \emph{{Dark
  matter from dark photons: a taxonomy of dark matter production}},
  \href{https://doi.org/10.1103/PhysRevD.100.095018}{\emph{Phys. Rev. D}
  {\bfseries 100} (2019) 095018}
  [\href{https://arxiv.org/abs/1908.09864}{{\ttfamily 1908.09864}}].

\bibitem{Belanger:2020npe}
G.~B\'elanger, C.~Delaunay, A.~Pukhov and B.~Zaldivar, \emph{{Dark matter
  abundance from the sequential freeze-in mechanism}},
  \href{https://doi.org/10.1103/PhysRevD.102.035017}{\emph{Phys. Rev. D}
  {\bfseries 102} (2020) 035017}
  [\href{https://arxiv.org/abs/2005.06294}{{\ttfamily 2005.06294}}].

\bibitem{Cirelli:2007xd}
M.~Cirelli, A.~Strumia and M.~Tamburini, \emph{{Cosmology and Astrophysics of
  Minimal Dark Matter}},
  \href{https://doi.org/10.1016/j.nuclphysb.2007.07.023}{\emph{Nucl. Phys. B}
  {\bfseries 787} (2007) 152}
  [\href{https://arxiv.org/abs/0706.4071}{{\ttfamily 0706.4071}}].

\bibitem{RevModPhys.53.43}
D.J.~Gross, R.D.~Pisarski and L.G.~Yaffe, \emph{Qcd and instantons at finite
  temperature}, \href{https://doi.org/10.1103/RevModPhys.53.43}{\emph{Rev. Mod.
  Phys.} {\bfseries 53} (1981) 43}.

\bibitem{Binder:2018znk}
T.~Binder, L.~Covi and K.~Mukaida, \emph{{Dark Matter Sommerfeld-enhanced
  annihilation and Bound-state decay at finite temperature}},
  \href{https://doi.org/10.1103/PhysRevD.98.115023}{\emph{Phys. Rev. D}
  {\bfseries 98} (2018) 115023}
  [\href{https://arxiv.org/abs/1808.06472}{{\ttfamily 1808.06472}}].

\bibitem{2022arXiv220903932B}
A.~{Bharucha}, F.~{Br{\"u}mmer}, N.~{Desai} and S.~{Mutzel}, \emph{{Axion-like
  particles as mediators for dark matter: beyond freeze-out}}, {\emph{arXiv
  e-prints} (2022) arXiv:2209.03932}
  [\href{https://arxiv.org/abs/2209.03932}{{\ttfamily 2209.03932}}].

\bibitem{Heeba:2018wtf}
S.~Heeba, F.~Kahlhoefer and P.~St\"ocker, \emph{{Freeze-in production of
  decaying dark matter in five steps}},
  \href{https://doi.org/10.1088/1475-7516/2018/11/048}{\emph{JCAP} {\bfseries
  11} (2018) 048} [\href{https://arxiv.org/abs/1809.04849}{{\ttfamily
  1809.04849}}].

\bibitem{Slatyer:2021qgc}
T.R.~Slatyer, \emph{{Les Houches Lectures on Indirect Detection of Dark
  Matter}},
  \href{https://doi.org/10.21468/SciPostPhysLectNotes.53}{\emph{SciPost Phys.
  Lect. Notes} {\bfseries 53} (2022) 1}
  [\href{https://arxiv.org/abs/2109.02696}{{\ttfamily 2109.02696}}].

\bibitem{Pospelov:2007mp}
M.~Pospelov, A.~Ritz and M.B.~Voloshin, \emph{{Secluded WIMP Dark Matter}},
  \href{https://doi.org/10.1016/j.physletb.2008.02.052}{\emph{Phys. Lett. B}
  {\bfseries 662} (2008) 53} [\href{https://arxiv.org/abs/0711.4866}{{\ttfamily
  0711.4866}}].

\bibitem{Fabbrichesi:2020wbt}
M.~Fabbrichesi, E.~Gabrielli and G.~Lanfranchi, \emph{{The Dark Photon}},
  \href{https://arxiv.org/abs/2005.01515}{{\ttfamily 2005.01515}}.

\bibitem{Agashe:2014yua}
K.~Agashe, Y.~Cui, L.~Necib and J.~Thaler, \emph{{(In)direct Detection of
  Boosted Dark Matter}},
  \href{https://doi.org/10.1088/1475-7516/2014/10/062}{\emph{JCAP} {\bfseries
  10} (2014) 062} [\href{https://arxiv.org/abs/1405.7370}{{\ttfamily
  1405.7370}}].

\bibitem{Jaeckel:2014qea}
J.~Jaeckel, J.~Redondo and A.~Ringwald, \emph{{3.55 keV hint for decaying
  axionlike particle dark matter}},
  \href{https://doi.org/10.1103/PhysRevD.89.103511}{\emph{Phys. Rev. D}
  {\bfseries 89} (2014) 103511}
  [\href{https://arxiv.org/abs/1402.7335}{{\ttfamily 1402.7335}}].

\bibitem{Backovic:2015soa}
M.~Backovi\'c, M.~Kr\"amer, F.~Maltoni, A.~Martini, K.~Mawatari and M.~Pellen,
  \emph{{Higher-order QCD predictions for dark matter production at the LHC in
  simplified models with s-channel mediators}},
  \href{https://doi.org/10.1140/epjc/s10052-015-3700-6}{\emph{Eur. Phys. J. C}
  {\bfseries 75} (2015) 482}
  [\href{https://arxiv.org/abs/1508.05327}{{\ttfamily 1508.05327}}].

\bibitem{Englert:2016joy}
C.~Englert, M.~McCullough and M.~Spannowsky, \emph{{S-Channel Dark Matter
  Simplified Models and Unitarity}},
  \href{https://doi.org/10.1016/j.dark.2016.09.002}{\emph{Phys. Dark Univ.}
  {\bfseries 14} (2016) 48} [\href{https://arxiv.org/abs/1604.07975}{{\ttfamily
  1604.07975}}].

\bibitem{Becker:2022iso}
M.~Becker, E.~Copello, J.~Harz, K.A.~Mohan and D.~Sengupta, \emph{{Impact of
  Sommerfeld effect and bound state formation in simplified t-channel dark
  matter models}}, \href{https://doi.org/10.1007/JHEP08(2022)145}{\emph{JHEP}
  {\bfseries 08} (2022) 145}
  [\href{https://arxiv.org/abs/2203.04326}{{\ttfamily 2203.04326}}].

\bibitem{Arina:2020tuw}
C.~Arina, B.~Fuks, L.~Mantani, H.~Mies, L.~Panizzi and J.~Salko, \emph{{Closing
  in on $t$-channel simplified dark matter models}},
  \href{https://doi.org/10.1016/j.physletb.2020.136038}{\emph{Phys. Lett. B}
  {\bfseries 813} (2021) 136038}
  [\href{https://arxiv.org/abs/2010.07559}{{\ttfamily 2010.07559}}].

\bibitem{Arina:2020udz}
C.~Arina, B.~Fuks and L.~Mantani, \emph{{A universal framework for t-channel
  dark matter models}},
  \href{https://doi.org/10.1140/epjc/s10052-020-7933-7}{\emph{Eur. Phys. J. C}
  {\bfseries 80} (2020) 409}
  [\href{https://arxiv.org/abs/2001.05024}{{\ttfamily 2001.05024}}].

\bibitem{Holdom1986TwoUA}
B.~Holdom, \emph{Two u(1)'s and epsilon charge shifts}, {\emph{Physics Letters
  B} {\bfseries 166} (1986) 196}.

\bibitem{Shtabovenko:2020gxv}
V.~Shtabovenko, R.~Mertig and F.~Orellana, \emph{{FeynCalc 9.3: New features
  and improvements}},
  \href{https://doi.org/10.1016/j.cpc.2020.107478}{\emph{Comput. Phys. Commun.}
  {\bfseries 256} (2020) 107478}
  [\href{https://arxiv.org/abs/2001.04407}{{\ttfamily 2001.04407}}].

\bibitem{Kahlhoefer:2015vua}
F.~Kahlhoefer, K.~Schmidt-Hoberg, J.~Kummer and S.~Sarkar, \emph{{On the
  interpretation of dark matter self-interactions in Abell 3827}},
  \href{https://doi.org/10.1093/mnrasl/slv088}{\emph{Mon. Not. Roy. Astron.
  Soc.} {\bfseries 452} (2015) L54}
  [\href{https://arxiv.org/abs/1504.06576}{{\ttfamily 1504.06576}}].

\bibitem{Tulin:2017ara}
S.~Tulin and H.-B.~Yu, \emph{{Dark Matter Self-interactions and Small Scale
  Structure}}, \href{https://doi.org/10.1016/j.physrep.2017.11.004}{\emph{Phys.
  Rept.} {\bfseries 730} (2018) 1}
  [\href{https://arxiv.org/abs/1705.02358}{{\ttfamily 1705.02358}}].

\end{thebibliography}\endgroup
\bibliographystyle{JHEP_mod}

\end{document}